\documentclass[printer]{aa}
\usepackage{graphics,epsfig,latexsym,longtable,lscape}

\def\ergsec{\hbox{erg s$^{-1}$}}
\def\ergcm{\hbox{erg cm$^{-2}$ s$^{-1}$}}

\def\cm-2{cm$^{-2}$}
\def\etal{et al. }
\def\HI{\hbox{H\,{\sc i}}}
\def\HII{\hbox{H\,{\sc ii}}}
\def\NH{N$_{\rm H}$}
\def\ein{{\it Einstein }}
\def\chandra{{\it Chandra}}

\begin{document}
 
   \thesaurus{3; 
              (09.10.1; 
               11.09.1 NGC 253; 
               11.19.2; 
               11.19.3; 
               13.25.2) 
             }
   \title{X-ray observations of the starburst galaxy NGC~253:}
   \subtitle{II. Extended emission from hot gas in the nuclear area, disk, 
                 and halo}
   \author{W.~Pietsch\inst{1} \and
           A.~Vogler\inst{2,1} \and
           U.~Klein\inst{3} \and
           H.~Zinnecker\inst{4}}
   \offprints{W.~Pietsch}
   \mail{wnp@mpe.mpg.de}

   \institute{Max Planck Institut f\"ur extraterrestrische Physik,
              Giessenbachstra\ss e, D-85740 Garching bei M\"unchen, Germany
   \and       CEA/Saclay, DAPNIA, Service d'Astrophysique,
              L'Ormes des Merisiers, B\^at. 709,
              F--91191 Gif-sur-Yvette, France
   \and       Radioastronomisches Institut der Universit\"at Bonn,
              Auf dem H\"ugel 71, D-53121 Bonn, Germany
   \and       Astrophysikalisches Institut Potsdam, 
              An der Sternwarte 16, D-14482 Potsdam, Germany
        }          
   \date{Received date; accepted date}
   \titlerunning{Diffuse X-ray emission from NGC~253} 
   \maketitle

   \begin{abstract}
Spatial and spectral analysis of deep ROSAT HRI
and PSPC observations of the near edge-on starburst galaxy NGC 253 reveal 
diffuse soft X-ray emission, which contributes 80\% to its total X-ray 
luminosity (L$_{\rm X} = 5\,10^{39}$ \ergsec, corrected for foreground 
absorption). The nuclear area, disk, and halo contribution to the luminosity
is about equal. The starburst nucleus itself is highly absorbed and not 
visible in the ROSAT band.

The emission from the nuclear area stems from a heavily absorbed source with an
extent of 250 pc (FWHM) about 100 pc above the nucleus along the SE minor axis
(``nuclear source", X34), 
and the ``X-ray plume". The nuclear source is best described as having a thermal 
bremsstrahlung spectrum with a temperature of T = 1.2 keV
(N$_{\rm H} = 3\,10^{21}$ cm$^{-2}$) and   
L$_{\rm X}^{\rm exgal} = 3\,10^{38}$ \ergsec\ 
(corrected for Galactic foreground absorption). 
The spectrum of the hollow-cone shaped plume (opening angle of 32\degr\ and 
extent of $\sim$ 700 pc along the SE minor axis) is best modeled by a 
composite of a thermal bremsstrahlung 
(N$_{\rm H} = 3\,10^{20}$cm$^{-2}$, T = 1.2 keV, L$_{\rm X}^{\rm exgal} = 
4.6\,10^{38}$ \ergsec) and a thin 
thermal plasma (Galactic foreground absorption, T = 0.33 keV, 
L$_{\rm X}^{\rm exgal} = 4\,10^{38}$ \ergsec). The diffuse nuclear
emission components trace interactions between the galactic super-wind emitted 
by the starburst nucleus, and the dense interstellar medium of the disk.

Diffuse emission from the disk is heavily absorbed and follows the spiral 
structure. It can be described by a thin thermal plasma spectrum ( T = 0.7 keV, 
intrinsic luminosity L$_{\rm X}^{\rm intr} = 1.2\,10^{39}$ \ergsec), 
and most likely reflects a mixture of sources (X-ray binaries, supernova 
remnants, and emission from \HII\ regions) and the hot interstellar medium. 
The surface brightness profile reveals a bright inner and a fainter outer 
component along the major axis with extents of $\pm$3.4 kpc and $\pm$7.5 kpc. 

We analysed the total halo emission separated into two geometrical areas; 
the ``corona" (scale height $\sim1$ kpc) and the ``outer halo".
The coronal emission (T = 0.2 keV,  
L$_{\rm X}^{\rm intr} = 7.8\,10^{38}$ \ergsec) is only detected from the 
near side of the disk (in the SE), emission from the back (in the NW) is 
shadowed by the intervening interstellar medium unambiguously determining the
orientation of NGC 253 in space. In the NW we see
the near edge of the disk is seen , but the far component of the halo, and vice 
versa in the SE. The emission in the outer halo 
can be traced to projected distances from the disk of 9 kpc, and shows a 
horn-like structure. Luminosities are higher (10 and $5\,10^{38}$ \ergsec, 
respectively) and spectra harder in the NW halo than in the SE. 
The emission in the corona and outer halo 
is most likely caused by a strong galactic wind emanating from the starburst 
nucleus. As an additional contribution to the coronal emission 
floating on the disk like a spectacle-glass, we propose
hot gas fueled from galactic fountains originating
within the boiling star-forming disk. 
A two temperature thermal plasma model
with temperatures of 0.13 and 0.62 keV or a thin 
thermal plasma model with temperature of 0.15 keV and Gaussian
components above $\sim$0.7 keV and Galactic foreground absorption 
are needed to arrive at acceptable fits for the NW halo. This
may be explained by starburst-driven super-winds or by effects of a 
non-equilibrium cooling function in a plasma expanding in fountains 
or winds. 

We compare our results to observations at other wavelengths and from other
galaxies. 
 
   \end{abstract}

\section{Introduction}
A hot gaseous component in the
interstellar medium with a temperature around 10$^6$ K is expected to originate
from SNRs in the disk of spiral galaxies. The medium 
might emerge via galactic fountains into the halo of these galaxies
(e.g. Spitzer 1956; Cox \& Smith 1974; Bregman 1980a,b;
Corbelli \& Salpeter 1988). Supernovae and winds from massive stars in a
central starburst might even drive a large-scale outflow that can shock heat
and accelerate ambient interstellar or circumgalactic gas in form of a
galactic super-wind (e.g. Heckman et al. 1990).
This component (e.g. Cox \& Reynolds 1987) was
detected by Snowden \etal (1994) in the plane and halo of the Milky Way
during the all sky survey and in pointed
observations performed by the R\"ontgen observatory  satellite (ROSAT). Hot interstellar gas in
the LMC originally detected by observations of the \ein observatory
(e.g. Wang \etal 1991) was confirmed in the `first light' observations 
with ROSAT (Tr\"umper \etal
1991). Outside the local group, \ein upper limits to the diffuse emission from
hot gas were determined for edge-on galaxies (Bregman \& Glassgold 
1982) and for the large face-on galaxy M101 (McCammon \& Sanders 1984).
For the starburst galaxies M82, NGC~253, and NGC~3628 however,
Watson \etal (1984), Fabbiano \& Trinchieri (1984), Fabbiano (1988) and
Fabbiano \etal (1990) resolved extended emission that was attributed to gaseous 
clouds ejected from the starburst 
nuclei, with temperatures in the $10^6$ K range.

Deep PSPC and HRI observations of the prototypical starburst galaxy 
NGC~253, due to its low Galactic foreground absorption and big optical extent, 
are ideal exploiting the ROSAT virtues 
(Tr\"umper 1983, Aschenbach 1988, Pfeffermann et al. 1987). The low 
Galactic foreground absorption (N$_{\rm H} = 1.3\,10^{20}$ cm$^{-2}$, 
Dickey \& Lockman 1990) allows soft X-rays 
from NGC~253 to
reach the detector with little attenuation. Pietsch (1992) reported first 
results based on some early ROSAT PSPC observations. 
Read et al. (1997) presented X-ray parameters 
of NGC~253 in a homogeneous analysis of archival ROSAT PSPC data of nearby 
spiral galaxies, and Dahlem
et al. (1998) investigated archival ROSAT and ASCA data in an X-ray 
mini-survey of nearby edge-on starburst galaxies including NGC 253.

   \begin{table}
      \caption{Parameters of NGC~253.}
         \label{galpar}
         \begin{flushleft}
         \begin{tabular}{lrr}
            \hline
            \noalign{\smallskip}
 & &Ref.  \\
            \noalign{\smallskip}
            \hline
            \noalign{\smallskip}
Type &  Sc & $^\ast$ \\
            \noalign{\smallskip}
Assumed distance & 2.58 Mpc  & 2 \\
& (hence 1$'\cor750$~pc) & \\
            \noalign{\smallskip}
Position of &  $\alpha_{2000}= 0^{\rm h}~47^{\rm m}~33\fs3$& 3\\
center (2000.0) & $\delta_{2000}=-25\degr 17\arcmin 18''$ & \\
            \noalign{\smallskip}
$D_{25}$ & 25\farcm 4 & 1 \\
            \noalign{\smallskip}
Corrected $D_{25}$ & 18\farcm 8 & 1 \\
            \noalign{\smallskip}
Axial ratio & 0.23 & 1 \\
            \noalign{\smallskip}
Position angle & 52\degr & 4 \\
            \noalign{\smallskip}
Inclination & 78.5\degr & 5 \\
            \noalign{\smallskip}
Galactic foreground $N_{\rm H}$ &1.3$\times10^{20}$~cm$^{-2}$ &6 \\
            \noalign{\smallskip}
            \hline
            \noalign{\smallskip}
         \end{tabular}
         \end{flushleft}
{
References: (1) Tully (1988);
(2) Puche \& Carignan (1988);
(3) Forbes et al. (1991);
(4) SIMBAD data base, operated at CDS, Strasbourg, France;
(5) Pence (1980)
(6) Dickey \& Lockman (1990)
}
   \end{table}
 
In this paper we present a detailed analysis of the 
diffuse X-ray emission of NGC~253, characterizing in detail the different
emission components of the area close to the nucleus (``nuclear source" and
``X-ray plume"), of the disk, and the halo hemispheres. 
The point source contributions were separated with the help of the 
point source catalog presented by Vogler \& Pietsch (1999, Paper I).
Galaxy parameters used throughout this paper are summarized in 
Table \ref{galpar}.
Our analysis is restricted to ROSAT data. Results of ASCA or
BeppoSAX are hampered by the limited spatial resolution of these instruments 
which does not allow the separation of the very different surface brightness 
components, present in NGC 253.
We compare our findings with results from other X-ray investigations 
(including ASCA and BeppoSAX) and
with other wavelengths and discuss the emission components in view of 
starburst driven super-wind models. 

\section{Observations and point source analysis}

   \begin{table}
      \caption{Full width at half maximum (FWHM) of the ROSAT telescope/PSPC
          detector point spread function (PSF) at 0.3 and 1 keV for different
          off-axis angles. Cut radii used to discriminate against contributions of 
          detected point sources to the diffuse emission of the galaxy as given
          in the text can be estimated from these values}
         \label{psf}
         \begin{flushleft}
         \begin{tabular}{lrrrrr}
            \hline
            \noalign{\smallskip}
Energy &\multicolumn{5}{c}{off-axis angle}  \\
(keV)  & 0\arcmin & 5\arcmin & 10\arcmin & 15\arcmin & 20\arcmin \\ 
            \noalign{\smallskip}
            \hline
            \noalign{\smallskip}
0.3 & 38\arcsec & 38\arcsec & 41\arcsec & 50\arcsec & 65\arcsec \\
1.0 & 24\arcsec & 24\arcsec & 29\arcsec & 40\arcsec & 58\arcsec \\
            \noalign{\smallskip}
            \hline
         \end{tabular}
         \end{flushleft}
   \end{table}
 
NGC~253 was observed with the ROSAT HRI and PSPC for 57.7~ks 
and 22.8~ks, respectively. 
For the analysis all available ROSAT observations of the PSPC and HRI have
been merged. 
Details of the observations and methods used to
derive the X-ray point source catalog are described in
Paper I.  
The analysis of extended emission components is based on
the corrected event files and images created for Paper I' and 
on additional procedures 
using the ESO-MIDAS/EXSAS (ESO-MIDAS 1997, Zimmermann et al. 1997) software 
package. While the higher resolution of the HRI detector 
allows us to resolve the bright nuclear and X-ray plume area, the lower instrumental 
background, the energy resolution and the high collecting area in the
0.1--0.4 keV band makes the PSPC best suited for studying large-scale 
diffuse emission in the disk and halo of the galaxy. 

Separating diffuse emission components from point source contributions is an
ambitious task keeping in mind the limited spatial 
resolution of the ROSAT PSPC detector. To lose as little area as
possible, point sources contained in the source catalog of Paper I were cut 
out with radii optimized to the point spread function (PSF) of the 
telescope/detector at the relevant energy (see Table \ref{psf}). 
Details on rejected sources and corresponding cut 
radii are mentioned below when describing the methods used to characterize
diffuse emission. To keep the uncertainty low,  
the derived diffuse fluxes and luminosities are not extrapolated for the 
full area in the standard procedure. To improve the estimates for the nuclear 
area we use the higher resolution data of the HRI (see Sect. 3.2.1).

The point source catalog of Paper I contains 73 sources in the
NGC~253 field, 32 of which are associated with the disk of the galaxy. 
Though 27 of these sources are detected with the HRI (some being resolvable
with the PSPC), the remaining 5 PSPC-only detected sources are most likely 
due to fluctuations within the diffuse X-ray emission (see Figs. 2 and 4 of
Paper I). 
The area around the nucleus is resolved into a bright, mildly absorbed
point source (X33, most likely a black hole X-ray binary) and a more highly 
absorbed, slightly 
extended nuclear source embedded in bright diffuse X-ray emission from the 
X-ray plume (already detected in \ein observations, see Fabbiano \& Trinchieri
1984, Fig. 2). The halo of NGC~253 is filled with diffuse, 
filamentary X-ray emission (cf. Fig. \ref{four_in_one}). While four 
of the sources detected within 
this region are most likely background active galactic nuclei (AGN, 
as suggested by time variability arguments and proposed optical identification),
two PSPC-only detected sources (X24 and X27 of Paper I) 
are most
likely spurious detections caused by local enhancements in the diffuse
emission. X10, another PSPC-only detected extended 
source in the NW halo, might be a background cluster of galaxies, 
as it shows a harder spectrum than the surrounding diffuse emission. 

\section{Analysis of diffuse emission and results}
Several methods are used to characterize the diffuse X-ray emission 
components of NGC~253. Iso-intensity contour maps exhibit the point
sources and diffuse emission with high resolution for the HRI. The maps 
are split up in several energy bands, 
however with lower spatial resolution for the PSPC. The images
in different PSPC energy bands can be combined to a ``true" colour
image of the galaxy. We further investigate the emission components
with respect to spatial distribution (box profiles along major and minor axis
and azimuthal profiles for different energy bands and the X-ray plume region) and
spectral behavior (e.g. integrated spectra of nuclear component, X-ray plume, disk,
and halo, temperature profiles of the halo). 
In the following subsections we describe the procedures used for the analysis 
and we present a first discussion of results.

\subsection{Iso-intensity contour and "true" colour maps}
\begin{figure*}
  \resizebox{\hsize}{!}{\includegraphics[bb= 56 262 561 767,clip=]{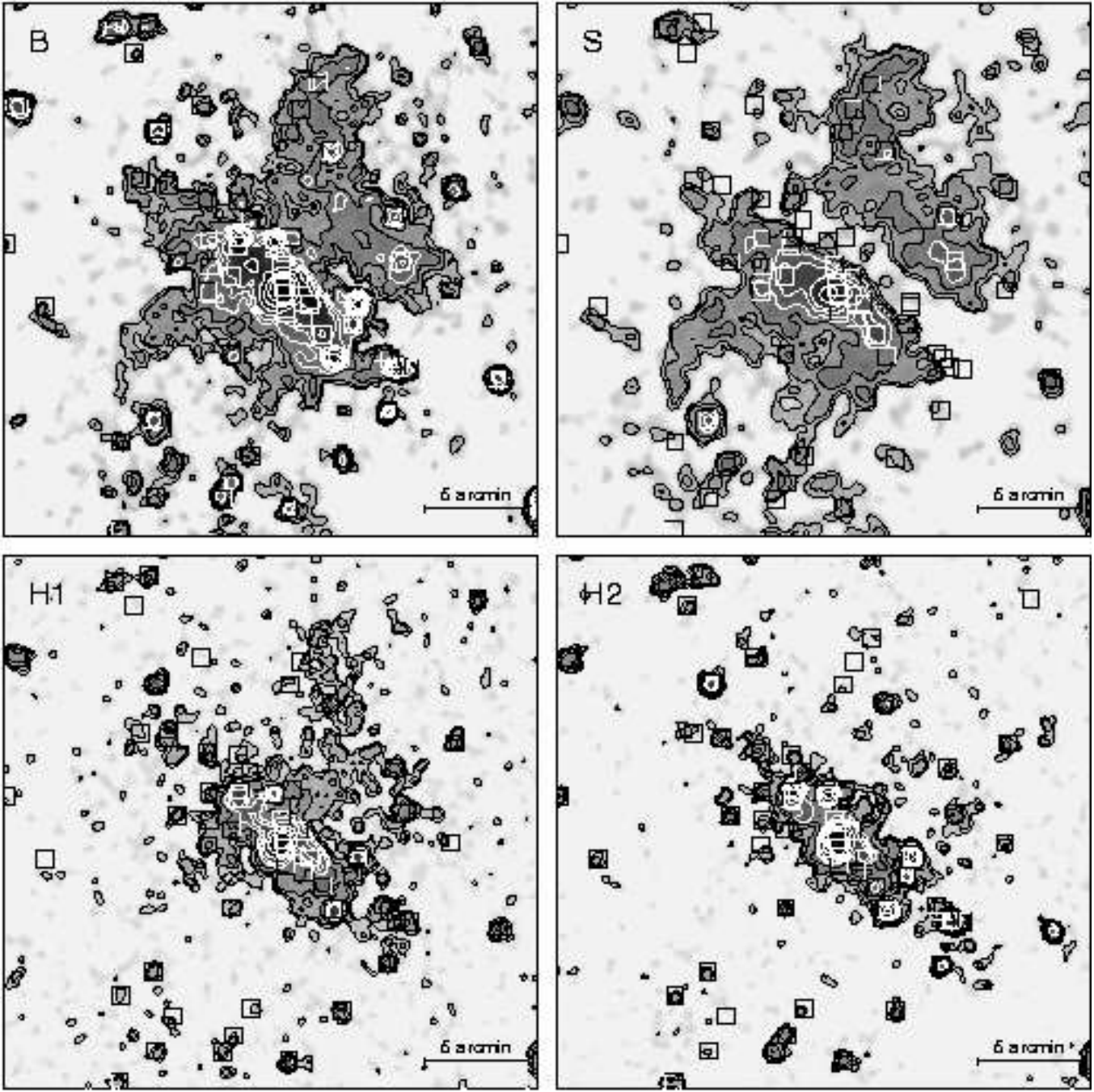}}
    \caption[]{
     Contour plots superimposed on smoothed grey-scale images of
     NGC~253 for broad (B), soft (S), hard1 (H1), and hard2 (H2) ROSAT
     PSPC bands.
     Broad and soft band contours are given units of $\sigma$ 
     ((323 and 240)$\times10^{-6}$\,cts~s$^{-1}$~arcmin$^{-2}$, respectively)
     above the background
     ((1590 and 1150)$\times10^{-6}$\,cts~s$^{-1}$~arcmin$^{-2}$, 
     respectively),
     hard band contours (due to the negligible background
     in these bands) in
     units of 1 photon accumulated per 28\arcsec\ diameter. One
     unit$\;=250\,10^{-6}$\,
     cts s$^{-1}$ arcmin$^{-2}$ for the hard bands.
     Contour levels are 3, 5, 8, 12, 17, 30, 60, 120
     units for all contour plots.
     Squares indicate sources contained in the NGC~253 source catalog 
     (Paper I)
     }
    \label{four_in_one}
\end{figure*}
\begin{figure*}
  \resizebox{12cm}{!}{\includegraphics[bb= 65 270 512 718,width=12cm,clip=]{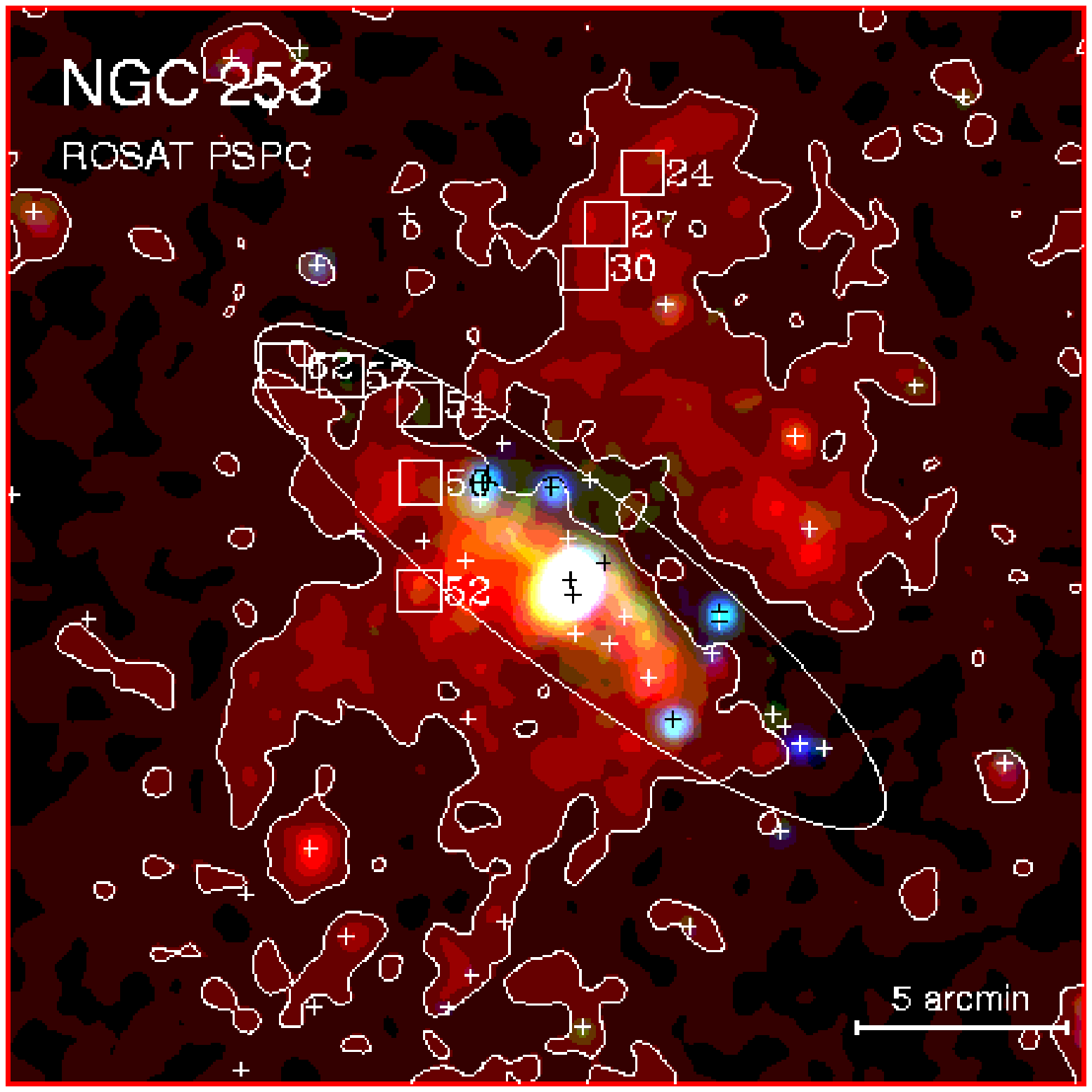}}
  \hfill
  \parbox[b]{55mm}{
    \caption[]{ROSAT PSPC ``true" colour image of NGC~253 constructed from
     the image of the soft band (0.1--0.4 keV) in red, hard1 band 
     (0.5--0.9 keV) in green, and hard2 band (1.0--2.0 keV) in blue. The
     optical size of NGC~253 is indicated by the inclination corrected
     D$_{25}$ ellipse. Sources from the point source list that were not 
     screened out for determining the diffuse emission are given as
     boxes with source numbers, while crosses mark the screened sources  
     }
    \label{rgb}}
\end{figure*}

The central area is best imaged with a contour plot from the HRI data. 
We have used an image with a bin size of 1\arcsec, selecting only raw 
channels 2--8 to reduce the background caused by UV emission and cosmic rays, 
and smoothed it with a Gaussian filter of 5\arcsec\ FWHM, corresponding
to the on-axis HRI PSF (cf. Fig. 3 of Paper I and Fig. \ref{halpha_hri}). 
The image shows two bright 
point-like sources, the central source X34,  and X33, 
embedded in a complex bright diffuse emission structure 
that is elongated in NW--SE direction with a maximal width of 45\arcsec\ 
(560 pc) protruding 60\arcsec\ (750 pc) from the nuclear source X34 to the SE, 
and 40\arcsec\ (500 pc) to the NW. While
the source $\sim$ 20\arcsec\ south of the nucleus (X33) is a time variable
point source (FWHM $\sim$5\arcsec, see Paper I), 
the source close to the nucleus (X34) is extended (FWHM $\sim$23\arcsec\ and
$\sim$19\arcsec, i.e. 290 pc and 240 pc in east/west and north/south 
directions, respectively),
with the intensity maximum offset by 8\farcs3 to the SE of the 
nucleus, possibly representing the hottest part of the gas out-flowing from the
nuclear starburst region (cf. Paper I). 

For the PSPC data, contour plots were obtained  from images constructed by
the superposition of sub-images with 5\arcsec\ pixel size
in the 8 standard bands (R1 to R8, cf. Snowden et al. 1994). All sub-images 
were corrected for
exposure, vignetting, and dead time, and were smoothed
with a Gaussian filter of a FWHM corresponding to the on-axis point
spread function (PSF) of that particular energy band.  The FWHM values
used a range from 52\arcsec\ to 24\arcsec. In Fig.~\ref{four_in_one} contour
plots are overlaid on grey scale images of the X-ray emission for the
broad (0.1--2.4 keV, upper left), soft (0.1--0.4 keV, upper right), hard1
(0.5--0.9 keV, lower left), and hard2 (0.9--2.0 keV, lower right) band, 
respectively. 

Visual inspection of the sub-images reveals several components of diffuse
emission: 

\begin{itemize}
\item Emission outside of the disk is absent in the hard2 map, and only
present in the NW hemisphere in the hard1.
\item The soft emission in both halo hemispheres suggests a filamentary
structure that cannot be fully resolved due to the limitations of the
ROSAT PSPC PSF.
\item The diffuse emission in the soft band in the SE shows a flat maximum 
(projected distance of about 45\arcsec\ SE of the NGC~253 nucleus) and 
seems to floats on the disk like a spectacle-glass 
(diameter of 16\arcmin\ $\cor$ 12 kpc), from which two ``horns" 
protrude (separation at the connection with the diffuse disk emission  
about 5\arcmin\ $\cor$ 3.7 kpc) reaching into the SE halo out to a projected 
distance of more than 7\arcmin\ (5.2 kpc), being turned inward. 
The horn emanating from the northern half of the disk is more pronounced.
\item The soft emission in the NW (extent 12\arcmin\ $\cor$ 9 kpc parallel 
to the disk and 8\farcm8 $\cor$ 6.6 kpc in perpendicular direction) is 
separated from the emission in the SE by $\sim$ 1\farcm2 $\cor$ 900 pc and 
also exhibits an inward turned horn-like structure.
Again, the horn emanating from the northern half of the disk is more pronounced.
\item The emission in the hard bands shows -- besides point sources -- a strong 
extended filamentary component from the inner disk that protrudes from a
hardly resolved bright nucleus and traces the inner spiral arms. In addition,
there is a less structured diffuse component of the disk.   
From inspection of the sub-bands R6 (0.9--1.3 keV) and R7 (1.3--2.0 keV) 
we find that the diffuse components vanish in the R7 band.
\item Both hard bands reveal emission protruding into
the NW halo along the minor axis. While in the hard2 band this component can 
only be traced for about 1\arcmin\ $\cor$ 750 pc 
(mainly in the area with highly reduced 
emission in the soft band), it extends along the northern
horn as far as the soft emission in the hard1 band.  
\end{itemize}

This complex structure manifests itself in the true colour image (Fig.~\ref{rgb})
built up from the PSPC soft (red), hard1 (green) and hard2 (blue) 
band images displayed in Fig.~\ref{four_in_one}. 
The output image is made up of ``pointers" to a special colour look-up table,
representing 216 colour blends spanning the possible combinations of 
6 intensity levels for each of the basic colours red, 
green and blue. With this method soft sources appear in red, with a ``red"
intensity proportional to the intensity in the soft image, harder sources in 
green and blue. The point sources in the disk area are
hard, the greenish colour of the diffuse disk emission suggests a hotter
or more strongly absorbed gaseous component compared to that in the halo. 
The NW rim of the disk is bluer than
the SE part which is most probably due to the fact that in the NW one mainly
detects hot gas absorbed in the disk while the emission in the SE is a
superposition of the softer halo emission in the near halo hemisphere 
and the disk component from the SE. 
The greenish colour connecting to the red NW halo 
again suggests emission from highly 
absorbed hot gas. Due to the cut values used, the nuclear 
area is over-exposed. 

\subsection{Separating diffuse emission components}
To quantify the findings of the last subsection, we now derive spatial 
distributions, spectra and luminosities for the individual diffuse components.
Simple power law (POWL), thermal bremsstrahlung (THBR), 
or thin thermal plasma (THPL) models (Raymond \& 
Smith 1977 and updates) and combinations are fitted to the different 
components as detailed below. Fits using MEKAL thin thermal plasma 
models (Mewe et al. 1985 and updates) did not give significantly 
different results. For the thin thermal plasma model we assume 
cosmic abundances as the starburst should have enriched the interstellar material
with processed material. The statistics of the data and the restricted
energy range of the ROSAT PSPC prohibit more detailed modeling.

\subsubsection{Diffuse emission from nuclear area and X-ray plume}
\begin{figure}
  \resizebox{\hsize}{!}{\includegraphics[clip=]{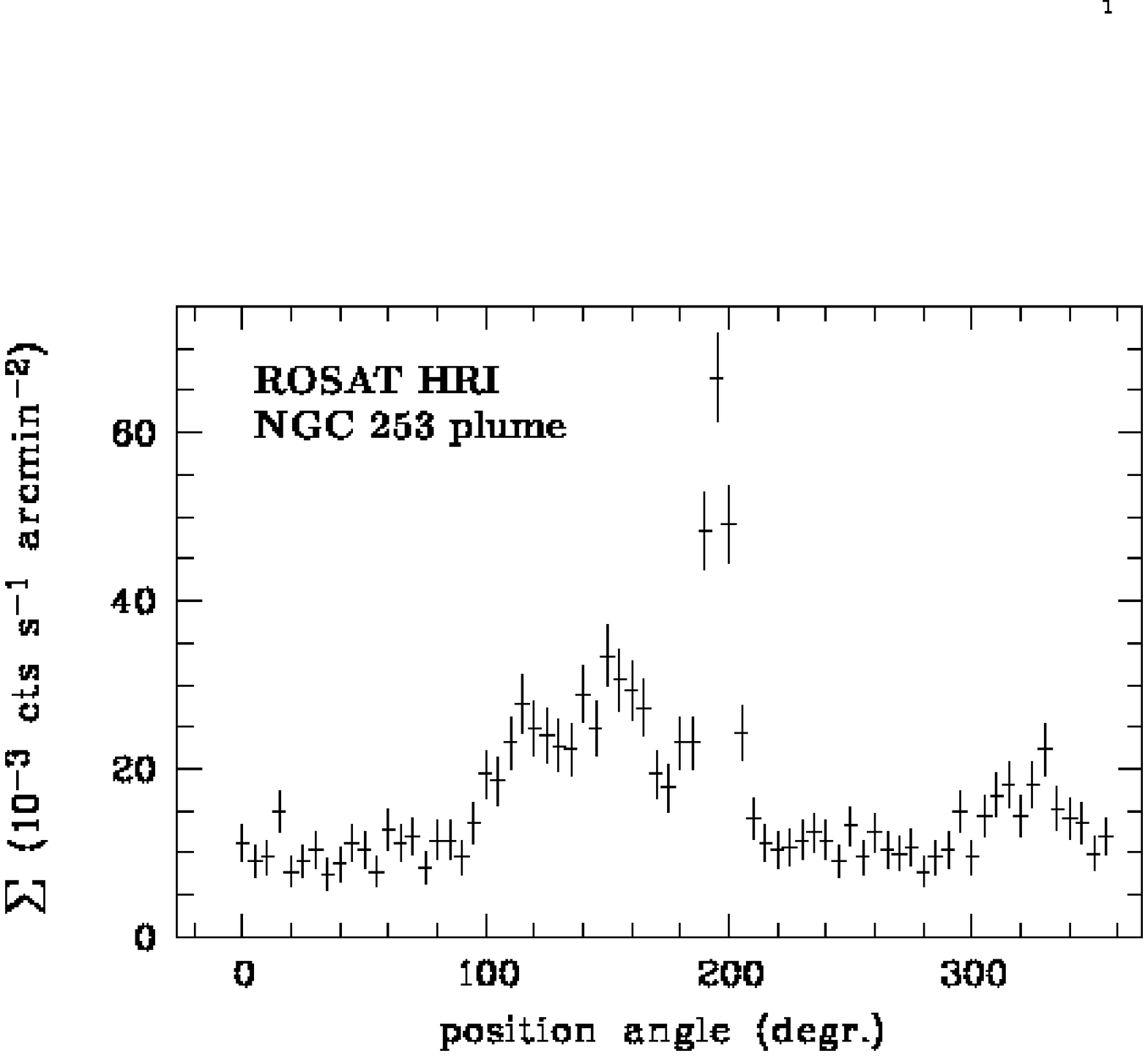}}
  \caption[]{
   Azimuthal distribution of the surface brightness of the 
   nuclear area and X-ray plume of NGC~253.
   ROSAT HRI counts are integrated  
   over sectors of 5\degr\ in an annulus with 15\arcsec\ and 1\arcmin\ 
   inner and outer radii, respectively (to exclude the central source). 
   0\degr\ is north
   and angles increase from north to east
   }
   \label{azim_hri}
\end{figure}

  \begin{table*}
      \caption{Source and background regions
       (and covered area), extracted net counts used in the subsequent
        spectral analysis, surface brightness, and hardness
        ratios are given for different areas of NGC 253. 
        Background was determined from four
        box regions of $5\arcmin\times5\arcmin$ aligned with 
        the axes, with their inner corners positioned at
        $\pm7\farcm5$ from the minor axis
        and $\pm2\farcm5$ from the major axis. Local background for
        the nuclear source (X34) was determined from the adjacent ring,
        with an outer radius of 30\arcsec\, 
        and screening the X33 sector.
        Point sources in all areas were screened with a radius 
        equivalent to the FWHM of the PSF (see text). This reduced the
        area for the disk, halo NW and SE integration by 17\%, 2\%, and 5\%,
        respectively}
         \label{spec-tab}
         \begin{flushleft}
         \begin{tabular}{llrrrrr}
            \hline
            \noalign{\smallskip}
Spectrum & Source region & Area  & Net counts & Surface brightness
       & HR1               & HR2         \\
       &  &(arcmin$^2$)  & (cts)       & (10$^{-3}~$cts s$^{-1}$~~~ & & \\
       &  &              &             &  arcmin$^{-2}$)& & \\
            \noalign{\smallskip}
            \hline
            \noalign{\smallskip}
{\it X34} & circle with radius of 20\arcsec\ around   
&0.28 &1008$\pm$32 & 158$\pm$5 & 0.92$\pm$0.01 & 0.35$\pm$0.03   \\
& nucleus, X33 sector screened  \\
            \noalign{\smallskip}
{\it X34\_local} & as above, but local background 
&0.28 &628$\pm$38 & 98$\pm$6 & 1.00$\pm$0.03 & 0.43$\pm$0.06   \\
            \noalign{\smallskip}
{\it X-ray plume}& ring sector around nucleus with
& 0.70 & 553$\pm$25& 34.6$\pm$1.6 & 0.49$\pm$0.04 & -0.12$\pm$0.05 \\
& inner and outer radii of 20\arcsec\ and\\
& 67\farcs5 opening angle 90\degr\ along SE \\
& minor axis, X33 screened with \\
& radius 26\arcsec \\
            \noalign{\smallskip}
{\it disk} & box enclosing the $D_{25}^c$ ellipse, &64.3&
2690$\pm$83 & 1.83$\pm$0.06 &0.23$\pm$0.03 &-0.10$\pm$0.03 \\
&nuclear area screened  \\
&with radius 67\farcs5 \\
            \noalign{\smallskip}
{\it halo NW} & box $15\arcmin\times10\arcmin$ adjacent to {\it disk} 
&147.2& 2590$\pm$110 & 0.77$\pm$0.03 &-0.33$\pm$0.04 &-0.31$\pm$0.06 \\
& along NW minor axis\\
            \noalign{\smallskip}
{\it halo SE} & box $15\arcmin\times10\arcmin$ adjacent to {\it disk} 
&141.9& 1437$\pm$103 & 0.44$\pm$0.03 &-0.75$\pm$0.06 &-0.39$\pm$0.23 \\
& along SE minor axis\\
            \noalign{\smallskip}
            \hline
          \end{tabular}
         \end{flushleft}
   \end{table*}

   \begin{table*}
         \caption{Spectral results in different regions described in 
                  Table \ref{spec-tab}. 
                  Errors ($1\sigma$) are only given in the case 
                  $\chi^2/\nu \le 2$}
         \label{spec}
         \begin{flushleft}
         \begin{tabular}{llrrrrrrrrr}
            \hline
            \noalign{\smallskip}
Spectrum& Model$^{~\ast}$ &$\chi^2/\nu$ &DOF
           & N$_{\rm H}$ & Index & T
           & f$_{\rm x}^{\rm ~exgal~ \S}$ & f$_{\rm x}^{\rm ~intr~ \S}$ 
           & L$_{\rm x}^{\rm exgal~ \S}$ & L$_{\rm x}^{\rm intr~ \S}$   \\
       \noalign{\smallskip}
           & & & &(10$^{20}$ cm$^{-2}$) & &(keV)  
           & \multicolumn{2}{c}{(10$^{-13}$\,\ergcm)}
           & \multicolumn{2}{c}{(10$^{38}$\,\ergsec)} \\
            \noalign{\smallskip}
            \hline
            \noalign{\smallskip}
{\it X34} &POWL&  1.53 &31 &$31.6_{-10.8}^{+19.5}$
               & $2.9_{-0.6}^{+1.1}$ & &5.6 &70 &4.5 &56 \\
            \noalign{\smallskip}
(S/N$\gid$5)  &THBR& 1.53 &31& $21.3_{-12.4}^{+9.9}$ & & $1.1_{-0.4}^{+2.3}$
                   &5.6 & 17 &4.5 &13 \\
            \noalign{\smallskip}
              &THPL& 2.78&31& $114$ & & $0.44$
                   &(4.5) &(130) &(3.6) &(110) \\
            \noalign{\smallskip}
              &THBR&1.09&30& $30.5_{-7.2}^{+10.0}$ & & $1.20^{\rm fix}$
                    &4.8 &16 &3.8 &13 \\
              &+THPL& & &$1.3^{\rm fix}$ & & $0.44_{-0.14}^{+0.58}$
                     &0.8 &0.8 &0.7 &0.7 \\
            \noalign{\smallskip}
            \noalign{\smallskip}
            \noalign{\smallskip}
{\it X34\_local} &POWL& 0.93 &12& $40.7_{-24.6}^{+33.7}$
                   & $2.9_{-1.3}^{+1.6}$ &  &3.8 &54 &3.0 &43\\
            \noalign{\smallskip}
(S/N$\gid$5)  &THBR&0.94&12& $29.3_{-15.9}^{+22.0}$ & & $1.2_{-0.6}^{+5.1}$
                   &3.7 &13 &3.0 &10 \\
            \noalign{\smallskip}
              &THPL& 1.22&12& $130_{-17}^{+22}$ & & $0.45_{-0.17}^{+0.22}$
                   &3.2 &120 &2.6 &96 \\
            \noalign{\smallskip}
            \noalign{\smallskip}
            \noalign{\smallskip}
{\it X-ray plume} &POWL& 1.67&17& $7.3_{-1.2}^{+1.4}$
                 & $3.2_{-0.3}^{+0.4}$ &  &3.2 &22 &2.6 &17\\
            \noalign{\smallskip}
(S/N$\gid$5)  &THBR&1.64&17& $4.7_{-0.8}^{+0.8}$ & & $0.52_{-0.09}^{+0.12}$
                   &3.0 &6.2 &2.4 &4.9\\
            \noalign{\smallskip}
              &THPL&4.4&18& $1.3^{\rm fix}$ & & $0.33$ 
                   &(2.2) &(2.2) &(1.8) &(1.8)\\
            \noalign{\smallskip}
              &THBR&1.37&16& $3.0_{-0.9}^{+1.8}$ & & $1.20^{\rm fix}$
                    &1.5 &2.3 &1.2 &1.8 \\
              &+THPL& & &$1.3^{\rm fix}$ & & $0.33_{-0.07}^{+0.13}$
                     &1.1 &1.1 &0.9 &0.9\\
            \noalign{\smallskip}
            \noalign{\smallskip}
            \noalign{\smallskip}
{\it disk} &POWL& 1.46&44& $5.7^{+0.9}_{-0.8}$ & $3.1_{-0.2}^{+0.3}$
              &  &14.8 &75 &11.9 &60 \\
            \noalign{\smallskip}
(S/N$\gid$5)  &THBR& 1.31&44& $3.2_{-0.5}^{+0.5}$ & & $0.56^{+0.08}_{-0.07}$  
                 &13.8 &23 &11.0 &18 \\
            \noalign{\smallskip}
              &THPL&5.8&45& $1.3^{\rm fix}$ & & $0.30$
                   &(10.1) &(10.1) &(8.1) &(8.1) \\
            \noalign{\smallskip}
              &THPL1&1.07&42&$95^{+23}_{-35}$ & & $0.69_{-0.29}^{+0.30}$
                     &4.2 &15.3 &3.4 &12.2\\
              &+THPL2& & & $1.3^{\rm fix}$ & & $0.20_{-0.02}^{+0.02}$
                    &9.8 &9.8 &7.8 &7.8 \\
            \noalign{\smallskip}
            \noalign{\smallskip}
            \noalign{\smallskip}
{\it halo NW} &POWL& 1.50&18& $4.2^{+1.5}_{-1.3}$ & $3.6_{-0.5}^{+0.5}$
                   &  &12.3 &75 &9.8 &60 \\
            \noalign{\smallskip}
(S/N$\gid$5)  &THBR& 1.14&18& $2.0_{-0.7}^{+0.7}$ & & $0.37^{+0.09}_{-0.07}$  
                   &12.3 &18 &9.8 &14 \\
            \noalign{\smallskip}
              &THPL& 3.8&19& $1.3^{\rm fix}$ & & $0.19$
                   &(12.7) &(12.7) &(10.1) &(10.1)\\
            \noalign{\smallskip}
              &THPL1&1.18&17& $1.3^{\rm fix}$ & & $0.13^{+0.02}_{-0.02}$
                   &9.1 &9.1 &7.2 &7.2 \\
              &+THPL2& & & $1.3^{\rm fix}$ & & $0.62^{+0.24}_{-0.21}$
                   &3.2 &3.2 &2.5 &2.5 \\
            \noalign{\smallskip}
            \noalign{\smallskip}
            \noalign{\smallskip}
{\it halo SE} &POWL& 2.1&9& $1.3^{\rm fix}$ & $3.2$
                   &  &(9.2) &(9.2) &(7.3) &(7.3) \\
            \noalign{\smallskip}
(S/N$\gid$4)  &THBR& 2.2&9& $1.3^{\rm fix}$ &
                   & $0.24$  &(7.0) &(7.0) &(5.6) &(5.6) \\
            \noalign{\smallskip}
              &THPL& 2.9&9 & $1.3^{\rm fix}$ & & $0.12$
                   &(6.2) &(6.2) &(5.0) &(5.0) \\
            \noalign{\smallskip}
              &THPL1&2.6&7& $1.3^{\rm fix}$ & & $0.10$
                   &(6.1) &(6.1) &(4.9) &(4.9) \\
              &+THPL2& & & $1.3^{\rm fix}$ & & $0.52$
                   &(0.8) &(0.8) &(0.6) &(0.6) \\
            \noalign{\smallskip}
            \hline
          \end{tabular}
         \end{flushleft}
\[
\begin{array}{lp{0.95\linewidth}}
$$^{\rm \ast}$$ & POWL: power law, THBR: thermal bremsstrahlung,
THPL: thin thermal plasma with cosmic abundances (Raymond \& Smith 1977 and updates), \\
$$^{\rm \S}$$ & for 0.1--2.4~keV band; $^{\rm exgal} \cor$ correction for
Galactic foreground absorption (1.3 10$^{20}$\ cm$^{-2}$); 
$^{\rm intr} \cor$ correction for no absorption \\
\end{array}
\]
   \end{table*}

As demonstrated in Paper I, 
the nuclear source (X34) peaks at $\alpha(2000.0) = 00^h47^m33\fs40, 
\delta(2000.0) = -25\degr17^m23\farcs1$, with a 90\% error radius (including
systematics) of $2\farcs5$, and that it 
is clearly extended. Outside a radius of
$\sim$ 15\arcsec\ there is a continuous transition to the diffuse
emission along the minor axis (cf. Fig.\ 3 of Paper I). 
The angular distribution of the 
emission components resolved with the HRI with respect 
to the position of the nucleus can be viewed in the
azimuthal plot of the surface brightness in the central area 
(radius 1\arcmin, Fig. \ref{azim_hri}), integrated over sectors of  
5\degr. In order to avoid the bright emission, the nuclear source
 was cut out with a radius of 
15\arcsec. Above a background level of 
$25.5\pm0.8$ counts per sector (determined in sectors 0\degr--90\degr\
and 210\degr--300\degr) $\cor \sim 10^{-2}$ cts s$^{-1}$ arcmin$^{-2}$ , 
three excesses can be identified:
\begin{enumerate}
\item
A bright peak ($393\pm24$ integrated counts above background) 
at 195\degr\ represents the point source X33. The 
number of counts extracted for X33 by this method is consistent with the 
results reported in Sect. 3.2 of Paper I ($359\pm21$ cts) derived 
with a different method.
\item
A broad, slightly asymmetric double-peak structure  (FWHM $\sim$75\degr, 
$546\pm34$ excess counts in sector 100\degr--175\degr) centered 
at position angle 145\degr\  is due to the extended emission to 
the SE along the minor axis. The two peaks at position angles 
$\sim$120\degr\ (FWHM of $\sim$35\degr, $238\pm21$ excess counts in sector 
100\degr--135\degr) and $\sim$150\degr\ (FWHM of $\sim$45\degr, 
$308\pm24$ excess 
counts in sector 140\degr--175\degr) may reflect 
a cone-like emission structure along the minor axis of NGC~253 
(position angle 142\degr), with an opening angle of $\sim$30\degr\  
originating from the galaxy nucleus.  
\item
A broad peak (FWHM $\sim$35\degr, $128\pm20$ excess counts in sector 
305\degr--345\degr) centered
at position angle $\sim$325\degr\  is due to the extended emission to 
the NW along the minor axis and pointing directly opposite 
to component two. According to Fig. 3 of Paper I 
the emission is less extended than that originating from the nuclear
area. This may be explained by higher
absorption in this direction (see discussion below). 
\end{enumerate}

To further characterize the diffuse nuclear emission components, 
we made use of
the spectral capabilities of the PSPC. This investigation, however, is  
hampered by the larger PSF of the PSPC compared to the HRI. Especially 
in the soft energy band (FWHM of PSF $\sim$ 40\arcsec) the
emission components separated by the HRI are difficult to resolve,and 
therefore only crude spectral
parameters and luminosities can be derived. A better estimate
of the luminosities of the nuclear source (X34) and the X-ray plume 
can be obtained by converting count rates 
from the ROSAT HRI detector to fluxes using the 
PSPC-derived spectral parameters (see below).   Detailed results for
the point source X33 south of the nucleus were already discussed 
in Paper I. 

Here we present source spectra for the 
extended nuclear source (X34) and the X-ray plume. The 
spectrum of X34 was extracted with a cut diameter of 40\arcsec. To suppress
contributions from the nearby source X33, a 70\degr\ sector in the 
direction of the source was excluded from the integration. 
We have used both, the background from the field outside the galaxy, used for
all other spectral investigations, and a locally defined background. 
The spectrum of the
extended emission along the SE minor axis (X-ray plume) was extracted
from a sector with opening angle 90\degr\ outside the nucleus excluding
X33 (with a circle radius of 26\arcsec, corresponding
to twice the expected FWHM of the PSF of a hard source).
Table \ref{spec-tab} summarizes the regions for source and background
and gives net counts and hardness ratios HR1 and HR2 (see Paper I 
for definitions). The hardness ratios indicate a highly absorbed and rather 
steep spectrum for the nuclear source, while
the X-ray plume is less absorbed and much softer (cf. Pietsch et al.
1998). We then used simple absorbed spectral models to fit the data in
the different regions, as indicated in Table \ref{spec}. 
Reduced $\chi^2$, number of degrees of freedom (DOF), absorption column,
photon index for power law and temperature for thermal spectra are given.  
Raw spectra were rebinned to obtain at least the signal to
noise level per bin given in Col. 1 of Table \ref{spec}. We give fluxes and
luminosities outside the Galaxy 
(f$_{\rm x}^{\rm ~exgal}$ and L$_{\rm x}^{\rm ~exgal}$, measured N$_{\rm H}$ 
minus Galactic N$_{\rm H}$), as well as intrinsic to the source
(f$_{\rm x}^{\rm ~intr}$ and L$_{\rm x}^{\rm ~intr}$, i.e. absorption zero). 
Rough errors for the fluxes and luminosities in Table \ref{spec}
are indicated by the statistical errors of the net counts (see
Table \ref{spec-tab}). Additional uncertainties arise from 
the poor knowledge of the spectrum, as can be seen by
comparing the fluxes derived for different models.

When the field background is used, 
simple power law (POWL), thermal bremsstrahlung (THBR), 
or thin thermal plasma (THPL) spectra with cosmic abundances (Raymond \& 
Smith 1977 and updates) do not allow a proper 
description of the nuclear spectrum (X34). 
If, however, the spectrum is corrected for the local background,
the first two models fit the nuclear spectrum quite well, and also
a THPL approximation cannot be totally rejected. All
models indicate high absorption (N$_{\rm H}$ of a few times 10$^{21}$ to more
than 10$^{22}$ cm$^{-2}$) and rather soft spectra 
(photon index 2.9 or temperatures from 0.4 to 1.2 keV). 
This suggests that with the local background the soft extended emission above
the nuclear area is properly subtracted. We have therefore attempted to take
it into account by fitting a more complex model to the data derived using the
field background. We have assumed a two-component model composed of 
a THBR spectrum with the 
temperature fixed to 1.2 keV (the value determined using local background)
and a THPL spectrum using cosmic abundance with fixed Galactic absorption. 
This resulted in an acceptable
fit. For the THBR component flux and absorption are consistent with 
the values for the local background spectrum, and for the THPL component we 
derive a temperature of 0.44 keV.  

For the {\it X-ray plume}, POWL and THBR spectra yield $\chi^2/\nu \sim 1.7$ and
therefore are hardly acceptable, while a THPL with cosmic abundance 
clearly has to be rejected 
($\chi^2/\nu > 4$). Assuming a two-component model of a THBR with the 
temperature fixed to 1.2 keV - the value
fitted for the nucleus using local background - and additionally a THPL 
spectrum with cosmic abundance, 
fixed to Galactic absorption, leads to an acceptable fit. The THBR
component is much less absorbed than for the nucleus and the temperature of 
the THPL with $\sim$ 0.3 keV a bit cooler than for the nuclear spectrum. 

As mentioned above, the spatial resolution of the ROSAT HRI allows us to 
separately estimate the contribution from X34, the X-ray plume, 
and the point source X33.
From the numbers given in Paper I we determine ROSAT HRI count rates of 
($12.2\pm0.5$) and ($27.4\pm1.1$) $\,10^{-3}$ cts s$^{-1}$ for the
nuclear source (local background subtracted) and the X-ray plume, respectively. 
Using the X34\_local THBR model parameters from the PSPC, the 
count rates for the nuclear source give an intrinsic luminosity 
of $1.16\,10^{39}$ \ergsec, slightly higher than the PSPC value. 
For the X-ray plume,
the soft and hard component of the PSPC-based two-component model lead to
contributions to the HRI
count rates that are a factor of 1.3 higher for the THBR than for the 
THPL component. Also the soft component of the
X34 spectrum - that is even hotter than the X-ray plume component - has to be 
added to the HRI flux. Therefore we estimate that the THBR and THPL component 
contribute with a ratio of 3 to 4 and we assume a THPL temperature of  
0.4 keV (average of X34 and X-ray plume values). With these assumptions
we calculate intrinsic luminosities of 4.6 and 4.0 $\,10^{38}$ \ergsec\  
for the THBR and THPL components of the X-ray plume spectrum, respectively. These 
values are about a factor of four above those derived with the PSPC. The
difference is caused by the PSPC extraction strategy: We used only part
of the PSF area of X34 for the spectrum to reject contributions for X33. 
Also the rejection radii for X33 and X34  
strongly reduce the area for the X-ray plume spectrum. 
 
However, we have to place a caveat here. It is clear from the
structured HRI and energy-resolved PSPC images that more than two components 
contribute to the spectra. They originate from the nuclear area, X-ray plume, 
disk and halo and certainly are seen 
through different amounts of absorbing matter. Using a local background for the 
nuclear spectrum allows us to reduce the influence of the surrounding components.
Nevertheless, the assumed spectral models will be a simplification. 
The fact that a thermal bremsstrahlung or power law spectrum fit
best for the absorbed nuclear component and also for the absorbed component 
of the X-ray plume, should not be over-interpreted but may only reflect that several
thin thermal components of differing temperature add up to the spectrum. This
may be an artefact of the limited energy coverage and resolution of the 
ROSAT PSPC, and can hopefully be resolved with the next generation of X-ray 
instruments.   

\subsubsection{Diffuse emission from disk and halo}
\begin{figure*}
  \resizebox{\hsize}{!}{\includegraphics[bb= 55 264 569 553,clip=]{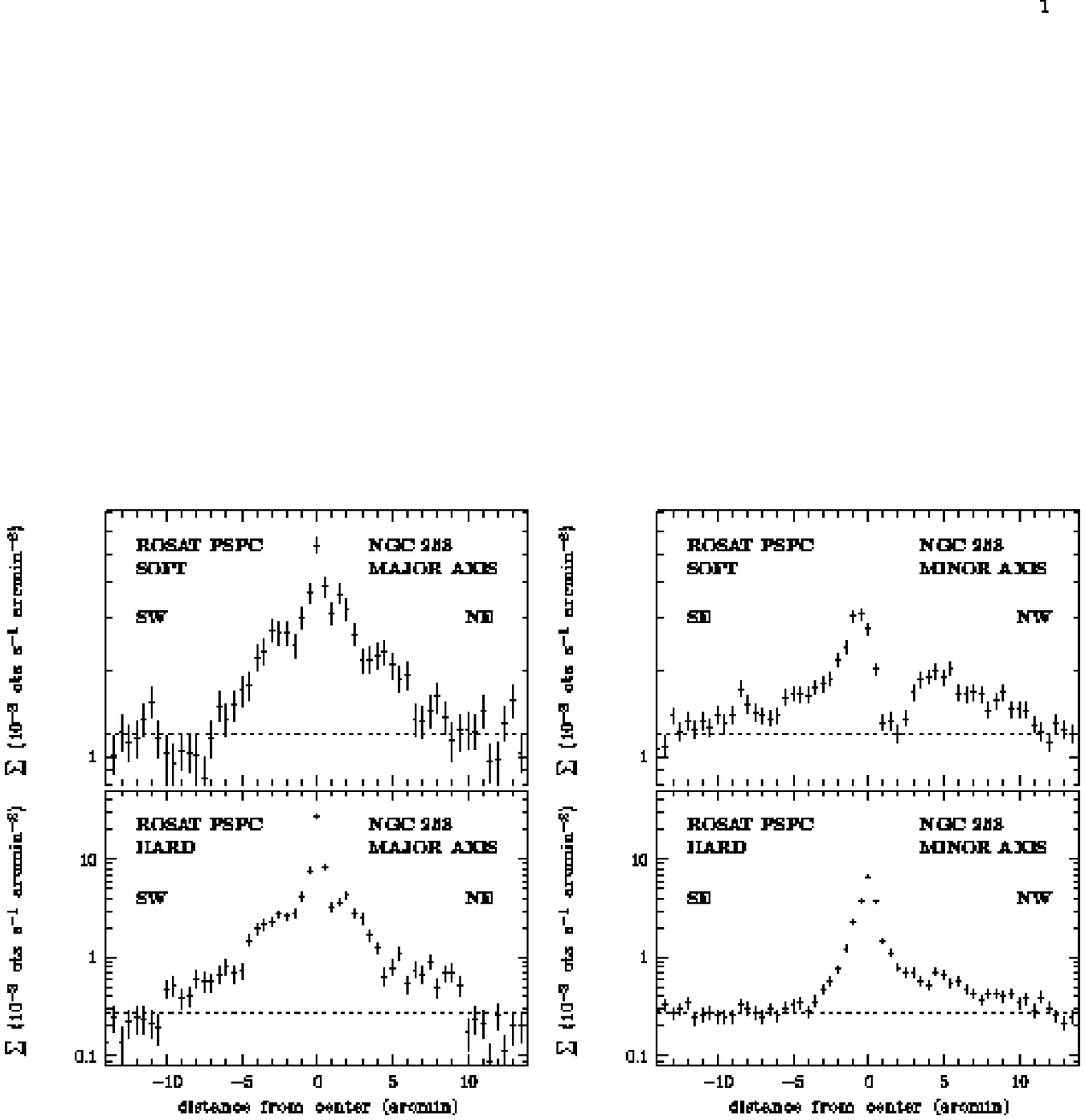}}
  \caption[]{
   Spatial distributions of surface brightness along the
   major (left) and minor (right) axes of NGC~253. 
   ROSAT PSPC soft (0.1--0.4 keV, top) and
   hard (0.5--2.0 keV, bottom) counts are
   integrated in boxes of $30\arcsec\times260\arcsec$\ along the major axis,
   covering the galaxy disk, and
   in boxes of $30\arcsec\times18\farcm8$\ along the minor axis,
   covering the galaxy halo region. Count rates are corrected for area, 
   exposure, dead-time, and vignetting. The boxes are
   centered at the distances given on the X-axis relative to the galaxy's 
   nucleus, with the SW disk region at left and NE to the right for the 
   major axis profile, and with the SE halo region at left and NW
   to the right for the minor axis profile.
   Contributions from point-like sources 
   are removed (see Sect. 3.2.2). Dotted lines indicate the
   background surface brightness determined in boxes of $>$12\arcmin\ distance
   from the nucleus using the minor axis profile
   }
   \label{profile_ax}
\end{figure*}

\begin{figure*}
  \resizebox{\hsize}{!}{\includegraphics[bb=55 264 569 553,clip=]{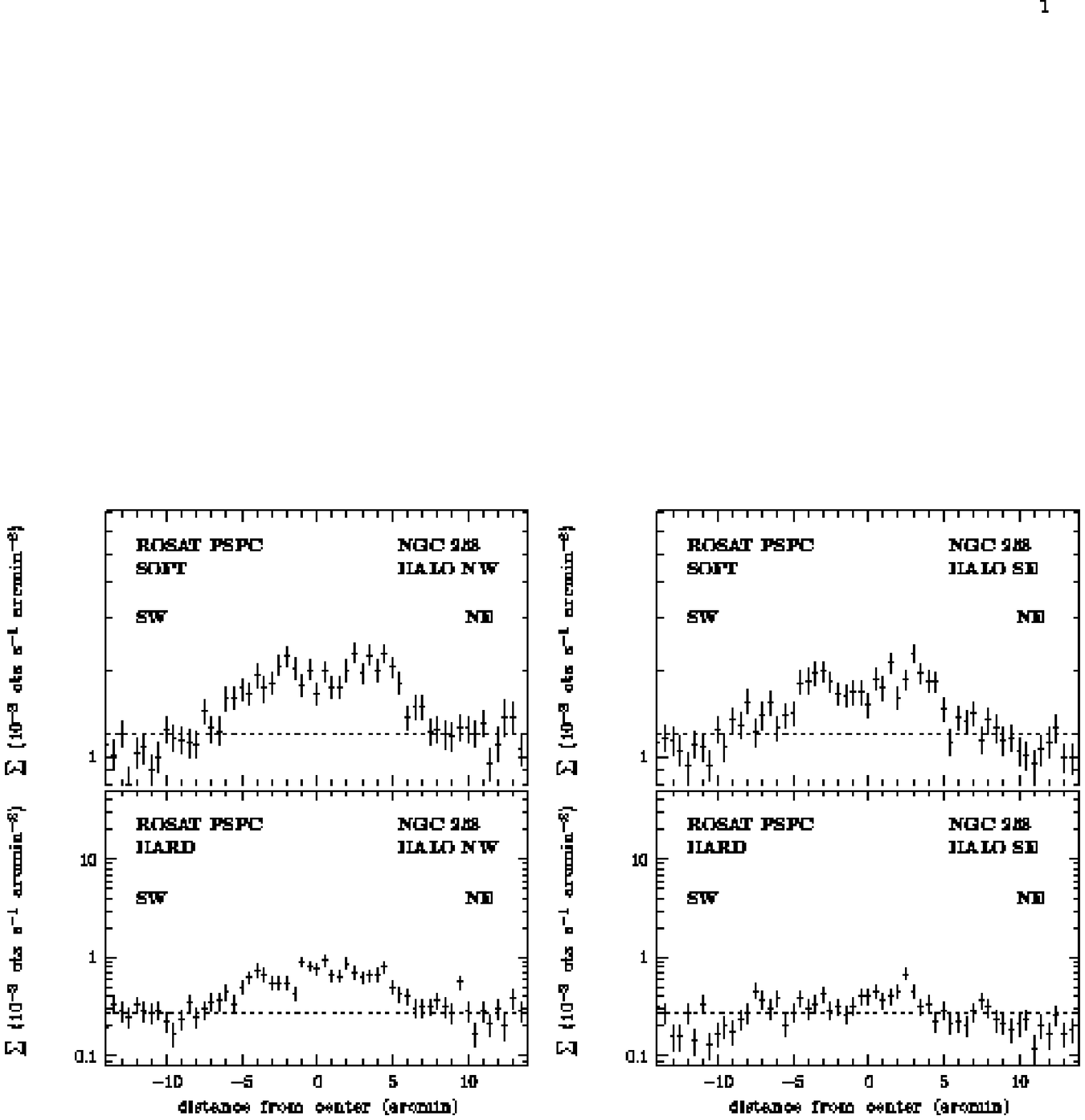}}
  \caption[]{
   Spatial distributions of surface brightness in the direction of
   major axis of NGC~253 offset to the NW (left) and SE (right). 
   ROSAT PSPC soft (0.1--0.4 keV, 
   top) and hard (0.5--2.0 keV, bottom) counts are integrated in boxes of 
   $30\arcsec\times480\arcsec$, choosing strips parallel to the major axis,
   covering the galaxy NW and SE halo region adjacent to the inclination 
   corrected D$_{25}$ ellipse, and corrected for area, 
   exposure, dead-time, and vignetting. The boxes are
   centered at the distances given on the
   X-axis relative to galaxy's minor axis, with the southern halo region to the 
   left. Contributions from point-like sources 
   are removed (see Sect. 3.2.2). Dotted lines indicate the
   background surface brightness as determined from Fig.~\ref{profile_ax}
   }
   \label{profile_off}
\end{figure*}

To investigate the diffuse emission from the disk and halo we make use of the 
nearly homogeneous sensitivity of the PSPC for low surface brightness features 
in the area within the ring of the window support structure (21\arcmin\ radius).
We examine box profiles along the galaxy axes, azimuthal profiles, and
spectra integrated over different areas. To suppress contributions from
point sources we follow a different strategy in creating profiles for the
individual energy bands and in extracting spectral files.

When creating profiles in the soft band, 
we cut out only sources outside the disk and X15 using a cut
radius of $1\times$\,FWHM of the PSF at 0.30 keV, since no source
from the point source catalog 
coincides with a point source in the soft band in the
disk area (within corrected D$_{25}$).  Furthermore, X24, X27, and X30 
(see Fig. \ref{rgb}) in the NW halo are not cut out as they are most 
likely ``spurious" sources (local enhancements in the diffuse emission, 
X24 and X27), or were not visible during the PSPC observations 
(X30). 

In the hard band, we cut out sources both in and outside of the disk. 
Outside we followed the same procedure as for the soft band, 
inside we cut out all sources
from the catalog, with the exception of the extended nuclear source (X33) and
the PSPC only detected sources X50, X51, X52, X57, and X62 (see Fig. \ref{rgb}),
most likely representing local enhancements in the diffuse emission in the 
disk. For the cut radius in the hard band we used $1\times$\,FWHM of the PSF at 
1.0 keV.

The spatial variation of the surface brightness distribution along the major 
axis of NGC~253 (Fig.~\ref{profile_ax}) was derived by integrating counts
in the soft and hard band in boxes of 30\arcsec\ width along this axis.
The boxes cover  $\pm130\arcsec$ perpendicular to the axis which 
corresponds to the maximum extent along the minor axis of the inclination 
corrected D$_{25}$ ellipse of NGC~253 (see Fig. \ref{rgb}). 
It is clear from Figs. \ref{four_in_one} and \ref{rgb} that by using this 
size of box perpendicular to the major axis of the edge-on galaxy,
one only traces emission components from the disk and/or
from the halo immediately above or below the disk. 
The neglected emission from the outer halo hemispheres is covered in the 
profile along the minor axis (boxes of 30\arcsec\ width along the minor axis 
and the corrected D$_{25}$ of 18\farcm8 in the direction of the major axis, 
Fig.~\ref{profile_ax}), and in profiles along axes parallel to the
major axis but offset to the SE and NW such that the covered strips are
adjacent to the major axis profile (boxes of $30\arcsec\times480\arcsec$, 
Fig. \ref{profile_off}). The background
surface brightness is determined from the minor axis count rates at distances
$>12\arcmin$. In this way, we utilize the high count statistics due to the 
biggest box size used for the different profiles 
and avoid a possible reduction of the extragalactic background due to 
absorption by an \HI\ disk extending further along the major axis 
than the galaxy's X-ray emission, as observed in some other galaxies (e.g. 
M101, Snowden \& Pietsch 1995; NGC 55, Barber et al. 1996; NGC 4559, 
Vogler et al. 1997).

Along the {\it major axis}, the soft- and hard-band profiles show a bright
nuclear component on top of a distribution that declines 
more or less symmetrically
with distance from the nucleus (Fig.~\ref{profile_ax}). 
In the hard band the distribution is composed of two distinct components
with extents of 4\farcm5 and 10\arcmin\ (3.4 kpc and 7.5 kpc) 
from the nucleus. By way of contrast the soft-band distribution 
is made up of just one component with 
an extent of $\sim$6\arcmin\ (4.5 kpc), which can be accounted for by  
absorption due to interstellar material within the disk. This is why we
detect more 
emission from a plane above the disk that is facing the line-of-sight
(SE disk corona). The
hard band profile is less effected by absorption and traces emission
from within, above and below the disk.

Along the {\it minor axis}, the soft and hard band profiles differ drastically 
(Fig.~\ref{profile_ax}). The hard-band profile is dominated by 
a bright core, positioned symmetrically about
the nucleus that can be traced out to 4\arcmin\ (3 kpc) to the SE, while to 
the NW is masked by a second component, with an
exponential decline out to $\sim$12\arcmin\ (9 kpc). 
In the soft band the maximum is shifted to the SE by $\sim$45\arcsec\ relative 
to the hard maximum, and it is not symmetric. To the SE, the bright core 
component drops within 3\arcmin\ (2.2 kpc) and then decreases more or less 
exponentially out to $\sim$12\arcmin\ (9 kpc) from the nucleus. In the NW, 
the profile first drops within 1\arcmin\ (750 pc) from the nucleus to 
just above the background level. Then, starting at a distance of 2\arcmin\ 
(1.5 kpc), the emission gradually recovers to a maximum at 
$\sim$4\arcmin\ (3 kpc), before it finally fades exponentially to background 
level at a similar distance from the nucleus as in the SE. 
Apart from the trough, the intensity in the NW is about 1.5 times that in the 
SE at similar offsets from the nucleus.
 
Parallel to the {\it major axis to the NW and SE} 
(Fig. \ref{profile_off}), the soft profiles show emission above the 
background out to projected distances from the nucleus of $\sim$6\arcmin\ 
(4.5 kpc). The profiles are double-peaked (broad maxima at offsets from
the nucleus of 
$\sim$3\farcm5 $\cor$ 2.6 kpc), reflecting the horn-like structure of the soft 
halo emission seen in Figs. \ref{four_in_one} and \ref{rgb}.
In the hard band, excess emission is clearly seen in the NW out to similar
distances as in the soft band. In the SE hard profile there is  
only a slight excess at a projected distance of $\sim$2\farcm5 
(1.9 kpc) from the nucleus to the NE, at the base of the corresponding 
horn-like structure in the images.

To further characterize the diffuse emission components from the disk and halo,
we analyzed PSPC spectra, integrated over areas summarized in Table
\ref{spec-tab}. In preparing the spectral files, we removed photons from 
source areas that were rejected from the soft and hard profiles (see
above) using cut radii of $1\times$\,FWHM of PSF at 0.3 or 1.0 keV (depending
on wether the sources were cut out in both bands or just in the hard one).
Results of fitting simple models are also given in Table \ref{spec}.

For the {\it disk spectrum}, POWL and THBR models yield 
$\chi^2/\nu \sim 1.4$ and therefore are hardly acceptable, 
while a single-temperature THPL with cosmic abundance 
clearly has to be rejected ($\chi^2/\nu > 5$).
However, a two-temperature cosmic abundance THPL model with the absorbing 
column of the softer component fixed to the
Galactic foreground results in an acceptable fit. In this model, one would
assign the unabsorbed 0.20 keV ($2.3\,10^6$ K) component to the hot gas above 
the disk and the absorbed $\sim$0.7 keV ($8\,10^6$ K) component
to the hot interstellar
medium within the disk. A even better approximation to reality would 
certainly require models
with more temperatures and absorbing columns. However, due to 
the limited statistics and spatial and spectral resolution of the ROSAT PSPC
data, more detailed modeling proves impossible.

\begin{figure}
  \resizebox{\hsize}{!}{\includegraphics[angle=-90]{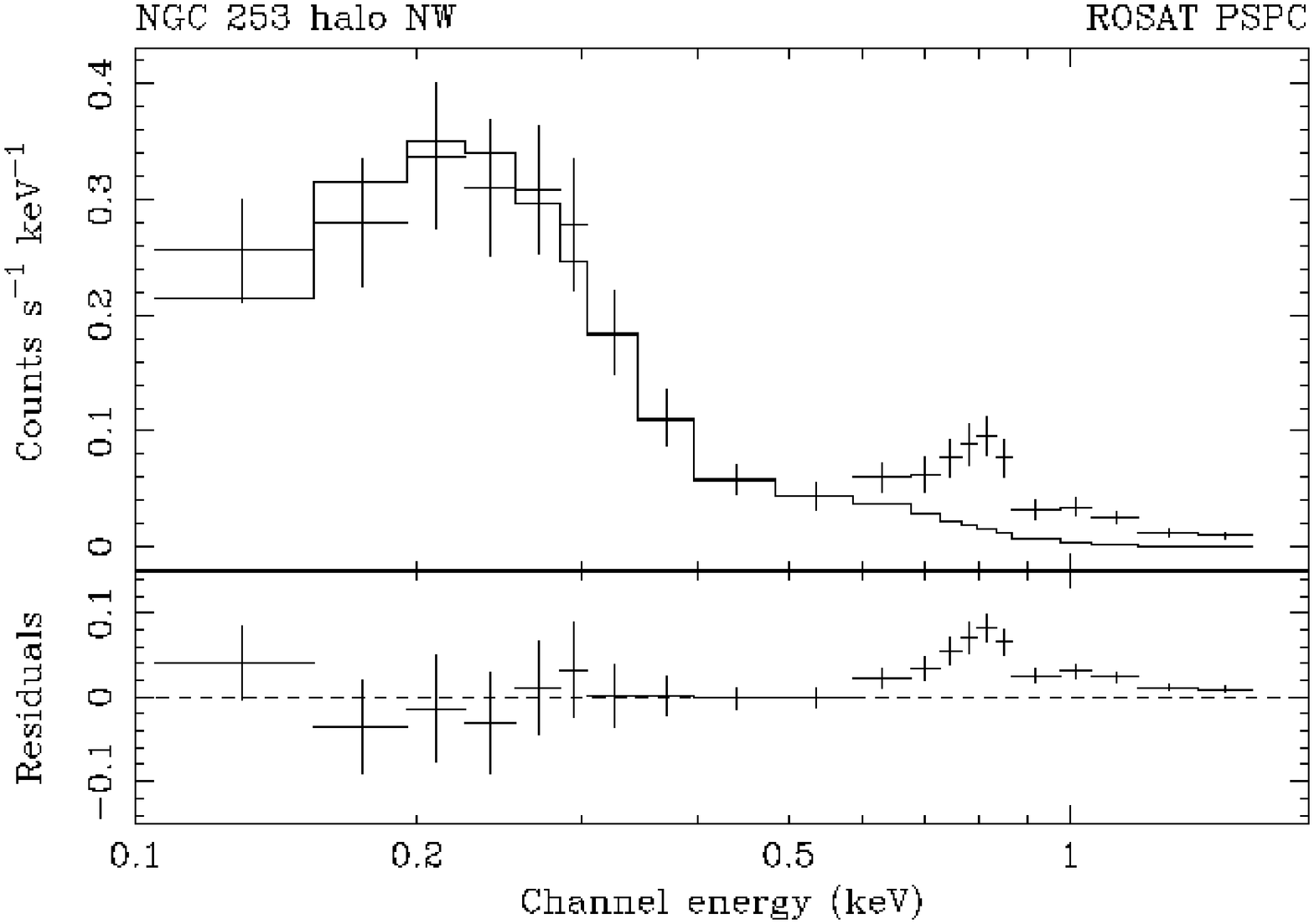}}
  \caption[]{
   PSPC spectrum of the NW halo of NGC 253 with a thin thermal
   plasma model (fitted only to the 0.1 -- 0.5 keV range, see text).
   The residuals are shown in the lower panel
   }
   \label{spectrum_h_nw}
\end{figure}

For the {\it NW halo spectrum}, POWL and THBR model approximations result in
barely acceptable fits not just from $\chi^2/\nu$ but also from 
systematics in the residuals between models and data. A cosmic abundance THPL model 
approximation is not acceptable ($\chi^2/\nu = 3.8$). If restricted to
the energy range 0.1 -- 0.5 keV, a cosmic abundance THPL model with absorption
fixed to the Galactic foreground and a temperature of 0.15 keV results in
an acceptable fit (see Fig. \ref{spectrum_h_nw}). Acceptable fits in the full
0.1 -- 2.4 keV range can be achieved
by adding additional emission components such as a second (hotter) thin thermal 
gas component or Gaussian lines at around 0.8 and 1.1 keV to such a spectrum. 

For the {\it SE halo spectrum} we obtain similar results. However, due to the poor 
photon statistics, the findings are less significant. Opposite to the NW halo, 
also a two temperature THPL model doesn't give an acceptable fit. Without acceptable fits
we cannot provide confidence regions for the fitted temperatures to compare them
to the NW halo hemisphere. This problem can be overcome in using the 
X-ray hardness ratios as a crude information on the spectral shape.  
For the SE halo, HR1 is smaller than for the NW halo ($\Delta $HR1 = -0.42$\pm$0.10) 
while the HR2 values coincide within the errors (see Table \ref{spec-tab}). 
This clearly indicates a significantly softer spectrum in the SE halo (see Fig.1 
in Pietsch et al. 1998). 

To make sure that the problems with fitting simple
models to the halo spectra were not caused by the special selection of the
halo regions we subdivided both halo areas into five boxes of 2\arcmin\ width
along the minor axis. Within the limits of the reduced statistics we
obtained similar results. Specifically the residuals at 0.7 keV and above
were present in all regions in the NW halo. This is consistent with the
detection of hard-band emission from the entire NW halo region that was 
already reported following the brightness profile analysis above. No 
significant temperature change could be established within the individual halo
hemispheres.

\section{Discussion}
In the following we will compare the diffuse emission components of 
NGC 253 presented in Sect. 3 to previously reported X-ray results for 
this galaxy. The individual
diffuse emission components are compared to observations of NGC 253 at other 
wavelength regimes, as well as to results from other spiral galaxies. 
We also derive parameters 
for the dense interstellar material in the disk of NGC 253 from its
apparent shadowing of X-ray emission in the NW halo. 
The findings are discussed in view of starburst and super-wind models.

\subsection{X-ray luminosity of NGC 253 emission components}
  \begin{table*}
         \caption{X-ray emission components of NGC 253: 
            ROSAT PSPC spectral results, count rates and luminosities 
            in 0.1--2.4 keV band (see Tables \ref{spec} and \ref{spec-tab} 
            and Sect. 4.1).
            Luminosities are given as observed (L$_{\rm x}^{\rm ~abs}$),
            corrected for Galactic foreground absorption 
            (L$_{\rm x}^{\rm exgal}$), and as intrinsic (corrected for zero
            absorption, L$_{\rm x}^{\rm intr}$).  
            Point source distributions from
            sources within the disk were taken from Paper I. It is obvious 
            that none of the components is dominant}
         \label{components}
         \begin{flushleft}
         \begin{tabular}{lllrrrrrr}
            \hline
            \noalign{\smallskip}
Component& &Spectral model$^{~\ast}$ & N$_{\rm H}$ & T
           & count rate & L$_{\rm x}^{\rm ~abs}$ 
           & L$_{\rm x}^{\rm exgal}$ & L$_{\rm x}^{\rm intr}$   \\
       \noalign{\smallskip}
           & & &(10$^{20}$ cm$^{-2}$) &(keV) &(10$^{-3}$cts s$^{-1}$)  
           & \multicolumn{3}{c}{(10$^{38}$\,\ergsec)} \\
            \noalign{\smallskip}
            \hline
            \noalign{\smallskip}
{\it X34\_local} & &THBR& 30 &1.2& 28 & 3.0 & 3.0&11 \\
            \noalign{\smallskip}
{\it X-ray plume} & &THBR& 3 &1.2& 46 & 4.1 & 4.6&6.9 \\
            & &+THPL&1.3&0.33& 47 & 3.3 & 4.0&4.0 \\
            \noalign{\smallskip}
{\it disk} & diffuse &THPL1& 95&0.7& 37 & 3.3 &3.4&12.2 \\
           &         &+THPL2 & 1.3&0.20& 78 &5.7&7.8 &7.8 \\
            \noalign{\smallskip}
{\it halo NW} & &THPL1& 1.3 &0.13& 93 & 6.0 & 7.2&7.2 \\
              & &+THPL2& 1.3&0.62& 22 & 2.1 & 2.5&2.5 \\
            \noalign{\smallskip}
{\it halo SE} & &THPL1& 1.3 &0.10& 55 & 1.9 & 4.9&4.9 \\
              & &+THPL2&1.3 &0.52&  6 & 0.5 & 0.6&0.6 \\
            \noalign{\smallskip}
            \hline
            \noalign{\smallskip}
{\it total} & diffuse & &  & & 412 & $\sim30$ & $\sim40$&$\sim60$ \\
            \noalign{\smallskip}
{\it disk} & point sources$^{~\rm \S}$& &  & & 94 & 8 & 10&(130) \\
            \noalign{\smallskip}
            \hline
            \noalign{\smallskip}
{\it total} & & &  & & 506 & $\sim38$ & $\sim50$&(190) \\
            \noalign{\smallskip}
            \hline
          \end{tabular}
         \end{flushleft}
\[
\begin{array}{lp{0.95\linewidth}}
$$^{\rm \ast}$$ & THBR: thermal bremsstrahlung,
THPL: thin thermal plasma\\
$$^{\rm \S}$$ & see Paper I \\
\end{array}
\]
   \end{table*}

As described in Paper I and in the previous section, the complex X-ray 
emission of NGC 253 can be separated into contributions from point sources
and diffuse emission. The diffuse emission originates from the nuclear area, 
the X-ray plume, the disk and from both halo hemispheres. In Table \ref{components}
we summarize the contributions of these components to the ROSAT PSPC count
rate and give luminosities in the ROSAT band, derived from the best fitting
spectra (see Table \ref{spec}). For the halo emission we use 
parameters derived for the two temperature thin thermal plasma models.
L$_{\rm x}^{\rm ~abs}$ is the absorbed luminosity as measured at 
the detector surface, L$_{\rm x}^{\rm exgal}$ the luminosity corrected for
Galactic foreground absorption of 1.3 10$^{20}$ cm$^{-2}$, 
and L$_{\rm x}^{\rm intr}$ the intrinsic 
luminosity assuming no absorption at all. While the results for the first
two are rather robust against model uncertainties due to the low Galactic
foreground absorption, the values for the intrinsic luminosity -- especially 
for the highly absorbed components -- are very sensitive to changes in the 
model parameters and therefore have to be taken with care. 

It is evident from the overall count rate budget in Table \ref{components},
that to first order, all components contribute similar amounts.
Due to the spatial resolution of the PSPC, it was possible to separately 
characterize each component's spectrum. 
As discussed below, most of the components
were already suggested in \ein observations. However, they could not be 
investigated in detail due to lack of statistics, and spatial and spectral 
resolution of the \ein IPC. In previous publications, ROSAT data for NGC 253 
were analyzed in 
investigations of samples of spiral and nearby edge-on starburst galaxies.
The special merits of ASCA and BeppoSAX observations of the galaxy are the
improved spectral resolution and range to higher energies covered. 
However, due to the comparatively low 
spatial resolution, the missing response below 0.5 keV and the limited 
statistics, not all components identified above can be spectrally resolved.
Within these limitations, the results of the other investigations support
ours.

A detailed analysis of the \ein HRI data of NGC 253 by Fabbiano \& Trinchieri
(1984) already revealed emission from several point sources in the
galaxy as well as
diffuse emission from the nucleus, along the minor axis to the SE and from
the inner disk. Assuming a 5 keV thermal bremsstrahlung spectrum with Galactic 
foreground absorption for the nucleus and disk and a 0.5 keV thermal plasma
spectrum for the X-ray plume, they derived luminosities in the 0.2--4 keV band 
(corrected to the NGC 253 distance of 2.58 Mpc used here) of 
(8, 3.3 and 10)$\times10^{38}$ \ergsec, respectively. 
Taking into account the difference in the model spectra and energy band, 
these values are consistent with our results. The analysis
of the \ein IPC observations of the galaxy by Fabbiano (1988) 
demonstrated that the emission profile from the inner disk of NGC 253 along 
the major axis closely follows the radio continuum emission. In addition
extended emission was found from the northern side of the galaxy and
attributed to gaseous clouds ejected from the starburst nucleus (luminosity
$1.1\,10^{39}$\ergsec). From the southern halo no emission was detected
with the \ein instruments. This, however, is not surprising since the  
collecting area of the \ein IPC was very low in the ROSAT PSPC soft band, 
where all emission for this component is detected. The limited spatial and 
spectral resolution as well as the lack of statistics hindered 
detailed spectral investigations with the \ein IPC.   

Earlier analysis of  ROSAT data did not discern the different emission 
components of NGC 253 in greater detail and therefore the results were still 
preliminary. The galaxy was e.g. analyzed as part of a 
sample of nearby spiral galaxies observed with the ROSAT PSPC (Read et al.
1997). For the sample galaxies, one- ore two-component spectra were fitted to
the emission as a whole and to the point sources and diffuse emission,
individually. For NGC 253, the authors found an integral
and diffuse luminosity of (8.1, 5.9)$\times10^{39}$ \ergsec\ (0.1-2 keV, 
corrected for a distance of 2.58 Mpc), respectively. The fraction of diffuse 
emission of 74\% for the luminosity escaping the galaxy compares 
well to $\sim 80$\% quoted in Table \ref{components}.
A one-component thermal plasma only poorly fits the diffuse emission, 
leading to an absorption compatible with the Galactic foreground, 
a temperature of 0.47 keV and heavy-element abundances of 0.02 solar.   
A two-component model, comprising a thin thermal plasma and an
absorbed, hard (10 keV) unresolved source component, improved the fit to 
the diffuse component and indicated that most ($\sim$90\%) of the 'diffuse' 
emission is truly cool (0.39 keV), low-metallicity (0.08 solar), diffuse gas, 
while the rest could be attributed to highly absorbed, hard sources. 
However, the fit is still not good ($\chi^2/\nu \sim 3.7$), 
and the authors argued that ``a much more
complex model, beyond the scope of this work, may be necessary to explain the 
halo emission from NGC 253 (and, indeed, other starbursts), as a large
temperature gradient is believed to exist within the halo of NGC 253".
In our detailed analysis, we could not establish a temperature gradient
within the halo as postulated by Read et al.. 
However we found differing temperatures for the 
diffuse emission components from the disk, the region immediately above the 
disk, and the individual halo hemispheres.

Ptak et al. (1997) report on the complex X-ray spectrum of NGC 253 as
measured with the ASCA instruments in the energy range 0.5--10 keV, which
shows strong emission lines from O, Ne, Fe, Mg, S, and Si above the continuum. 
Unfortunately, with ASCA it is not 
possible to spatially resolve point sources from diffuse emission or the 
different diffuse emission components. The integral spectrum can be fitted 
by two components, a ``soft" component described by  a temperature of
0.8 keV and an absorbed ``hard" component with a photon index of 2.0
or a temperature of 7 keV. They find that different models (with different
continua) yield absolute abundances that differ by more than an order of 
magnitude, while relative abundances are more robust and suggest an
under-abundance of Fe (inferred from the Fe-L complex) relative to 
$\alpha$-burning elements. The authors also try to derive element 
abundances from the individual line intensities and
argue that for the hard component, they have to be significantly sub-solar
(if thermal), or that there is a significant non-thermal or non-equilibrium
contribution. The ASCA spectral fit is confirmed by BeppoSAX observations
that, similar to ASCA, do not spatially resolve the components, though were, 
for the first time, able to detect the Fe K line at 6.7 keV 
(Persic et al. 1998). 

ROSAT and ASCA observations of NGC 253 are also included in an X-ray 
mini-survey of nearby edge-on starburst galaxies (Dahlem et al. 1998).
A ROSAT HRI image of the central area of NGC 253 is superimposed onto an 
H$\alpha$ image (see Pietsch (1994) for an overlay onto an optical one),
PSPC images in three energy bands, with and without point sources, indicate
diffuse emission up to the highest ROSAT band. Spectral results include a joint 
ROSAT and ASCA spectral study, and an investigation of individual areas 
with the ROSAT PSPC. Detailed discussions are deferred to a specific paper 
on NGC 253. In modeling the joint ROSAT ASCA spectrum, integrated over
the galaxy as a whole, Dahlem et al. find that the difficulty of measuring 
multiple absorbing columns causes the largest uncertainty. As an example,
they point out that it is impossible to measure the absolute or relative 
abundances or Fe with the integral spectrum, because N$_{\rm H}$ 
(which is at least
a few times 10$^{21}$ cm$^{-2}$ in the direction of the core) trades off 
directly against the Fe abundance -- stronger absorption for Fe L energies
than for Si and S lines at higher energies. They therefore conclude that for 
NGC 253 the absolute and relative abundances derived from the integrated ASCA 
and PSPC X-ray spectrum are not reliable indicators of the physical properties
of the gas. They also briefly discuss spectral modeling of compact sources,
core, disk and halo emission. However, the exact procedure is difficult to
reproduce. For the core, they do not separate the components of X33, X34, 
and the X-ray plume, and therefore, the results cannot be directly compared.
For the spectrum of the diffuse emission of the disk after point source 
subtraction, they reduce a local background, and they achieve an acceptable 
fit for a model that consists of an absorbed
power law and a thin thermal (0.25 keV) plasma with solar abundance (which 
corresponds to our two-component fit). The halo spectrum and three concentric 
sub-spectra are accumulated, averaging over both halo hemispheres.
Their halo spectrum is well fitted with a single temperature thin thermal 
plasma model with very low abundance. However, they prefer a two-temperature
plasma, with temperatures of 0.14 and 0.65 keV and solar abundance 
with a flux ratio of 4:1 and find no temperature dependence over the halo.
The hard spectral component is present in all 
three halo regions and not just close to the disk. 
These results are confirmed by our findings for the NW halo. Due to
the higher count rate, this hemisphere dominates the averaged spectra and 
probably masks in their analysis the significantly differing intensity and 
shape, we found in our analysis of the halo hemispheres.

\subsection{Emission from the area of the starburst nucleus and X-ray plume of 
            NGC~253}
\begin{figure*}
 \resizebox{12cm}{!}{\includegraphics[bb= 65 270 512 718,clip=]{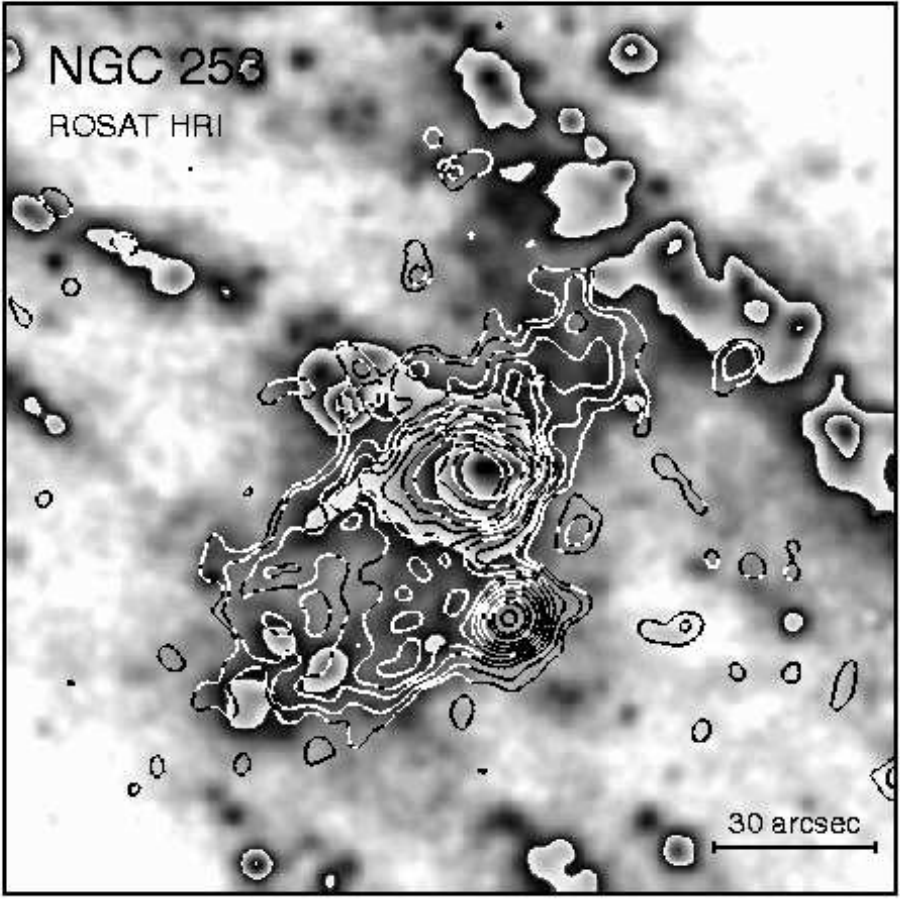}}
 \hfill 
 \parbox[b]{55mm}{
  \caption[]{
   Contour plot of the central emission region of
   NGC~253 for ROSAT HRI overlaid on a
   continuum-subtracted image of H$_{\alpha}$
   + [N\,{\sc ii} $\lambda\lambda$6548, 6583 line
   emission, kindly provided by H. Schulz (cf. Schulz \& Wegner 1992).
   The optical image is shown as a grey-scale representation, running 
   several times from bright to dark to make the full dynamical range
   of the data visible.
   X-ray contours are: 3, 4, 6, 8, 11, 15, 20, 27, 40, 50 and
   $70\times4.4\,10^{-3}$~cts~s$^{-1}$~arcmin$^{-2}$ 
   (same as in Fig. 3 of Paper I) }
   \label{halpha_hri}}
\end{figure*}

Following to Sect. 3.2.1 the X-ray emission from the nuclear
region of NGC 253 can be separated into at least three components: 
\begin{itemize}
\item {} the bright point source X33 about 25\arcsec\ south of the nucleus,
most likely a moderately absorbed black hole X-ray binary within NGC 253 
discussed in detail in Paper I
\item {} highly absorbed slightly extended emission SE 
of the nucleus (X34)
\item {} extended cone-like emission along the minor axis mainly to the 
SE of the nucleus -- the X-ray plume
\end{itemize}
To compare the nuclear emission components to the structures that are 
thought to trace the nuclear outflow of NGC~253, 
we superimposed in Fig. \ref{halpha_hri}
the HRI contours of the central area (contours as in Fig. 3 of Paper I) 
onto a continuum-subtracted H$_{\alpha}$ + [N\,{\sc ii} $\lambda\lambda
$6548, 6583 emission line image kindly provided by H. Schulz (cf.
Schulz \& Wegner 1992). 
X34 coincides with the central H$\alpha$ peak, and the
cone-like diffuse emission covers the "H$\alpha$ fan". Based on \ein HRI
data with lower statistical significance 
this fact was already noted by McCarthy et al.
(1987). They also pointed out that the H$\alpha$ bright emission regions to
the SE, near the end of the X-ray emission, are most likely ordinary \HII\  
regions in the disk of NGC 253. In the following we discuss the diffuse 
central X-ray components individually.

\subsubsection{Source X34: extended emission from the starburst nucleus 
of NGC~253?}
While Fabbiano \& Trinchieri (1984), using \ein HRI observations, identified
the extended source with the nucleus, the ROSAT HRI observations (cf. Paper I) 
demonstrate that it is offset by 5\farcs4 to
the east and by 6\farcs4 to the south from the position of the nucleus 
as defined by radio observations (Ulvestad \& Antonucci 1997, 
$\alpha(2000.0) = 00^h47^m33\fs10, \delta(2000.0) = -25\degr17^m17\farcs4$). 
This total offset of 8\farcs3 to the SE 
clearly exceeds the uncertainty in the X-ray position of 2\farcs5, 
including statistical and systematic errors. 
The maximum of the emission of X34 is also clearly 
separated from the radio-bright SNRs and \HII\ regions (e.g. Ulvestad
\&  Antonucci 1997), which are centered on the nucleus, and from the
bright near-infrared emission originating from dense dust clouds and
molecular material in the same region (Sams et al. 1994). We
therefore conclude that X34 is not emission originating from the 
nucleus of NGC 253 but represents the position, where emission from gas,
ejected along the minor axis, can penetrate into our line of sight through 
the dense absorbing interstellar medium surrounding the nucleus. 

The spectral results for X34 support this view, too. 
The N$_{\rm H}$ values of $(3-4)\times10^{21}$ cm$^{-2}$ derived for the 
X34 spectrum (subtracting a local background) are more than a factor of 10
below column densities of $\ga 6.3\,10^{22}$ cm$^{-2}$ expected for
emission from the NGC 253 nucleus. Such a high absorption is put forward
by the visual extinction of A$_{\rm v} > 35$ mag estimated for the nucleus by
Prada et al. (1999) based on Br$\gamma$ velocity curves along the SW side of
the major axis, and the conversion to N$_{\rm H}$ of 
$1.79\,10^{21}\times$A$_{\rm v}$ cm$^{-2}$ (Predehl \& Schmitt 1995).
A source with a power-law spectrum with photon index 2.9 as measured for X34
would be suppressed in the ROSAT 0.1--2.4 keV band 
by a factor of $>5000$ compared to zero absorption,
and still by a factor of $\sim 70$ compared to the N$_{\rm H}$ derived from
the X34 fit. If the spectrum of X34 was intrinsically flatter (e.g. power-law
of photon index 1.9), the attenuation in the ROSAT band due to absorption
might be reduced by just a factor of ten, but would still be substantial. 
Therefore, if hot gas is ejected from the nuclear area along the minor axis 
of the galaxy as a super-wind into the halo hemisphere facing us, soft 
band X-ray emission from this component will be heavily absorbed close to 
the nucleus and should become less absorbed along its path through the 
interstellar medium. 

The X-ray absorption is usually characterized by the hydrogen column density
N$_{\rm H}$  made up mostly of \HI\ and H$_2$. To calculate the absorption as 
function of X-ray energy one then uses the effective absorption cross section
per hydrogen atom as e.g. tabulated by Morrison \& McCammon (1983) based on
atomic cross sections and cosmic abundance. These abundances are derived in 
the solar neighborhood, and even within the Galaxy there is evidence for 
an radial abundance gradient for elements heavier than helium (Shaver et al.
1983). The X-ray absorption above 0.5 keV is primarily produced by heavier
elements, mostly by oxygen and iron and therefore dependent on the metal
abundance of the interstellar medium which may significantly differ from
solar in various regions of a starburst galaxy like NGC 253. Direct radio 
measurements of the \HI\ column in the direction of the NGC 253 nucleus
were hampered twofold, by the limiting beam resolution of $>$30\arcsec,  
and by \HI\ absorption. Therefore, the \HI\ column of 
$1.5\,10^{21}$ cm$^{-2}$ for the nucleus of NGC 253
obtained by the interpolation of the \HI\ maps of 
Puche et al. (1991) and Koribalski et al. (1995), can only be interpreted as
a lower limit. Mauersberger et al. (1996) investigate the distribution
of molecular gas in the direction of the NGC 253 nucleus and derive 
based on a conservatively low $N({\rm H}_2)/I_{\rm CO}$ value, a maximum
H$_2$ column density of $\sim 4\,10^{22}$ cm$^{-2}$ which (if half in the 
front and half in the back of the nucleus) contributes this amount to the
N$_{\rm H}$ towards the nucleus. The N$_{\rm H}$ determined in this way, 
nicely compares to the value based on the Br$\gamma$ velocity curves above.
Fits to broad band nuclear X-ray spectra with good statistics that will be 
derived with XMM-Newton or \chandra, should offer another method to determine the 
absorption column and possibly even abundances of the interstellar medium 
towards the nucleus of NGC 253.

With decreasing X-ray
energy, the maximum of the emission should be further separated from the 
position of the nucleus. This effect is indicated in
the PSPC images of the different energy bands in Fig. 1. However, one 
needs observations with high spatial resolution (as good as or better than
those provided by the ROSAT HRI) for several energy bands to trace the 
increasing absorption down to the nucleus and investigate its nature. As can
be seen from the numbers given above, an active nucleus with an intrinsic 
X-ray luminosity of more than a few times $10^{40}$ \ergsec, would still have 
been undetectable with ROSAT. 

The spatial displacement of X34 from the nucleus can be used to determine
the height above the plane of the galaxy from which the emission originates.
We assume a positional displacement along the minor axis and take into
account the inclination of NGC 253 (78.5\degr, Pence 1980). 
An offset of $8\farcs3\pm2\farcs5$ then 
transforms to a height of ($100\pm30$) pc above the nucleus. The extended 
nuclear source ($\sim 20$\arcsec\ FWHM) can be interpreted as hot gas being
part of a super-wind from the nuclear region. The THBR fit indicates a 
temperature of $1.3\,10^7$ K.
Its extent of $\sim$ 250 pc is similar in size to
the bright nuclear radio emission detected with the VLA at wavelengths ranging  
from 1.3 to 20 cm, where a large number of compact radio sources were revealed, 
embedded
within the diffuse radio structure (16\arcsec\ $\cor$ 200 pc along major axis, 
Ulvestad \& Antonucci 1997, and references therein), and more extended by a 
factor of two than the bright nuclear near infrared emission 
(Sams et al. 1994).

ASCA and BeppoSAX observations
(see Sect. 4.1) do not provide the resolution to spatially resolve the nucleus. 
However, if the hard component fitted to ASCA spectra is identified with 
emission from the nucleus, its intrinsic luminosity should not exceed 
$5\,10^{39}$ \ergsec\ (Ptak et al. 1997). As discussed before, this assumption
oversimplifies the situation since several sources with different absorptions 
contribute to the ASCA hard spectrum, the highly absorbed nuclear 
component representing just the one suffering the highest absorption. 
Its intrinsic luminosity may therefore be well in excess of 
the number given above
and does not really restrict the limits derived from the ROSAT observations.
The same arguments hold for the BeppoSAX observations, that show extended
emission best modeled by a hot thermal plasma ($kT \sim 6$ keV) with the
inclusion of an Fe K line and have been tentatively identified with
a starburst-driven galactic super-wind (Persic et al. 1998). There are, however,
also some similarities to the X-ray emission from the Galactic Ridge (Cappi
et al. 1999).  

The situation should change with the detectors on board the next generation of 
X-ray observatories, XMM-Newton and \chandra, that will allow observations at higher 
energies with good spatial resolution, strongly reducing the effects of 
absorption, and allowing the nucleus to become directly visible. With the help 
of X-ray variability studies it should then be possible to decide on the nature
of the radio-bright nuclear source which, according to Ulvestad \& Antonucci 
(1997), can still be either an AGN or a very compact supernova remnant. Recently,
time variability was detected in the highly absorbed hard X-ray component
of the nearby starburst galaxy M82 using ASCA data (Ptak \& Griffiths 1999, 
Matsumoto \& Tsuru 1999) which leads to the conclusion that M82 hosts
a low luminosity AGN. M82 shows X-ray emission from a strong galactic wind and 
the halo (e.g. Strickland et al. 1997). 
If indeed such a low-luminosity AGN were also confirmed for NGC 253, 
the  other nearby prototypical starburst galaxy with galactic wind 
and X-ray emission from the halo, this might strengthen the view of an 
intrinsic connection of these phenomena as proposed by Pietsch et al.
(1998), based on observations of the starburst galaxy NGC 3079, a galaxy 
which also hosts an active nucleus and a pronounced X-ray halo.

Forbes et al. (2000) present new HST data of the central region of NGC 253 
and compare it to other wavelengths. They find that the majority of 
optical/IR/mm sources are young star clusters which trace a $\sim50$ pc ring 
that defines the inner edge of a cold gas torus. In X-rays they identify the
extended nuclear ROSAT HRI emission with the nucleus and argue that not all of
the X-ray emission can be associated with the AGN, suggested from the radio, 
nor with an ultra-luminous supernova. Instead, they associate the 
emission with the out-flowing super-wind and find the size scale consistent with
the idea of collimation by the gas torus. This analysis in 
principle does conform with our above interpretation. The main difference
is that by careful positioning (see Paper I) we demonstrate 
that the emission in the ROSAT 
band is not originating from the nucleus but from a height above the plane
of NGC 253 where the extinction is already significantly reduced and 
transparent to radiation in the ROSAT band. At this height, the X-ray beam which
may be collimated in the inner disk by the cold gas torus as suggested above,
may already have widened by a factor $\ga 4$ as indicated from the extent 
reported in Sect. 3.1. 

\subsubsection{X-ray plume along SE minor axis}
Demoulin \& Burbidge (1970) were the first to suggest from spectrographic
observations of NGC 253 that gas might be flowing out of the nucleus and
outside the equatorial plane. In 1978, Ulrich confirmed the outflow along 
the minor axis with the help of velocity curves derived from measurements 
of optical lines using slit spectra across the
center of the galaxy with different orientations (to distances of 25\arcsec\  
from the nucleus). She explained the origin of the outflow in terms of a 
nuclear starburst. McCarthy et al. (1987) and Schulz \& Wegner (1992) used
long-slit spectra and narrow-band images and interpreted them as emission
from the surface of a kiloparsec-sized outflow cone driven by the starburst 
wind. The cone opening angle of 65 \degr\ and outflow speed along the cone of 
339 km s$^{-1}$ given by Heckman et al. (1990), compares well with
values determined by Schulz \& Wegner (50\degr\ and 390 km s$^{-1}$, 
respectively).  

The overlay of the ROSAT HRI contours onto the H$\alpha$ image 
(Fig. \ref{halpha_hri}) shows a rather close correspondence of the 
diffuse X-ray emission and the H$\alpha$ structures connected with the outflow.
The X-ray emission can be traced to a similar distance 
($\sim$ 1\arcmin ) from the nucleus to the SE as the weak H$\alpha$ 
and shows a clear cone-like structure with an opening angle of $\sim$30\degr, 
and brightened limbs in the ROSAT 
band (see Sect. 3.2.1) originating from the nucleus, which is also observed 
 in the H$\alpha$ images but less pronounced. The opening angle defined by
the maxima in Fig. \ref{azim_hri} is smaller 
than that based on optical radial velocities. This can be
explained by the near edge on viewing geometry, and as expected, the width of 
$\sim$75\degr\ of the emission structure as determined from Fig. \ref{azim_hri} 
in Sect. 3.2.1 compares more favorably.
The cone-like structure can be understood in terms
of models of galactic super-winds  
driven by a nuclear starburst (e.g. Tomisaka \& Ikeuchi 1988, Suchkov et al. 
1994). According to these models the cone surface may represent 
the {\it interaction zone} of the emerging wind with the interstellar medium
surrounding the nucleus. In H$\alpha$ and X-rays we do not
see the wind itself which is expected to be hotter than this medium 
and radiating predominantly at energies outside the ROSAT band. 
In this picture the relatively sharp cut-off 
of the X-ray emission to the SE can be understood as representing the
scale height of the dense interstellar medium in the galactic disk.

According to the models, a nuclear starburst should produce a 
bipolar flow along the minor axis. On the other hand, circumnuclear and disk 
material -- as discussed above -- will suppress
soft X-rays from the starburst wind emerging into the far, NW hemisphere. 
The HRI images still reveal some emission NW of the nucleus.
The morphology, however, is not that of a hollow-cone 
as in the SE. Therefore, we cannot decide whether this 
emission leaks through from the back of
NGC 253 through areas of reduced absorption (inter-arm region) in the 
galactic disk, or whether it originates in front of the disk. In principle
one could try to settle this question using the spectral capabilities of the 
PSPC, as we did for the nuclear source and X-ray plume. Unfortunately, the 
emission is not strong and extended enough and is too close to the bright 
central sources to allow this kind of analysis. 

However, the existence of a flow of gas into the NW halo hemisphere is 
demonstrated by the observation of an OH-plume pointing in this general 
direction, with a strong component to the north 
(Turner 1985). No OH emission (originating from molecular gas, radio 18 cm) 
is detected in the SE, adding to the
arguments that the outflow into the NW halo must even be stronger than  
that to the SE. This might be explained by a less strong blocking to the NW 
caused by inhomogeneities in the interstellar medium or an asymmetric 
position of the starburst with respect to the nucleus.
The intensities and spectra of the X-ray emission from the halo hemispheres
(see below) further support this picture.

The extended source above the nucleus (X34) and the X-ray plume can be directly 
compared to the X-ray emission from the nuclear super-bubble in the edge-on
starburst LINER/Seyfert 2 galaxy NGC 3079. There the
central component extends over 1.7 kpc with a luminosity of 
$7\,10^{39}$ \ergsec\ and is not only detected in H$\alpha$ images but
also in the 20 cm radio continuum (see Pietsch et al. 1998). Similar to
the NGC 253 X-ray plume, the central bright X-ray and H$\alpha$ emission 
is only seen in the east due to absorption in interstellar medium of this
galaxy. Another example of X-ray sources that may reflect emission from 
hot gas in super-bubbles emerging from galactic nuclei are reported by 
Read et al. (1995) for the merging galaxy pair NGC 4038/9 -- the Antennae.
There, the sources are offset by $\sim$1 kpc from the nuclei, are best fitted
by absorbed, low temperature plasma models indicating intrinsic luminosities
of a $(1-3)\times10^{40}$ \ergsec.  

\subsection{Diffuse emission from the NGC~253 disk and the flat SE halo
component immediately above (coronal emission)}
\begin{figure*}
  \resizebox{\hsize}{!}{\includegraphics[bb=57 262 569 594,clip=]{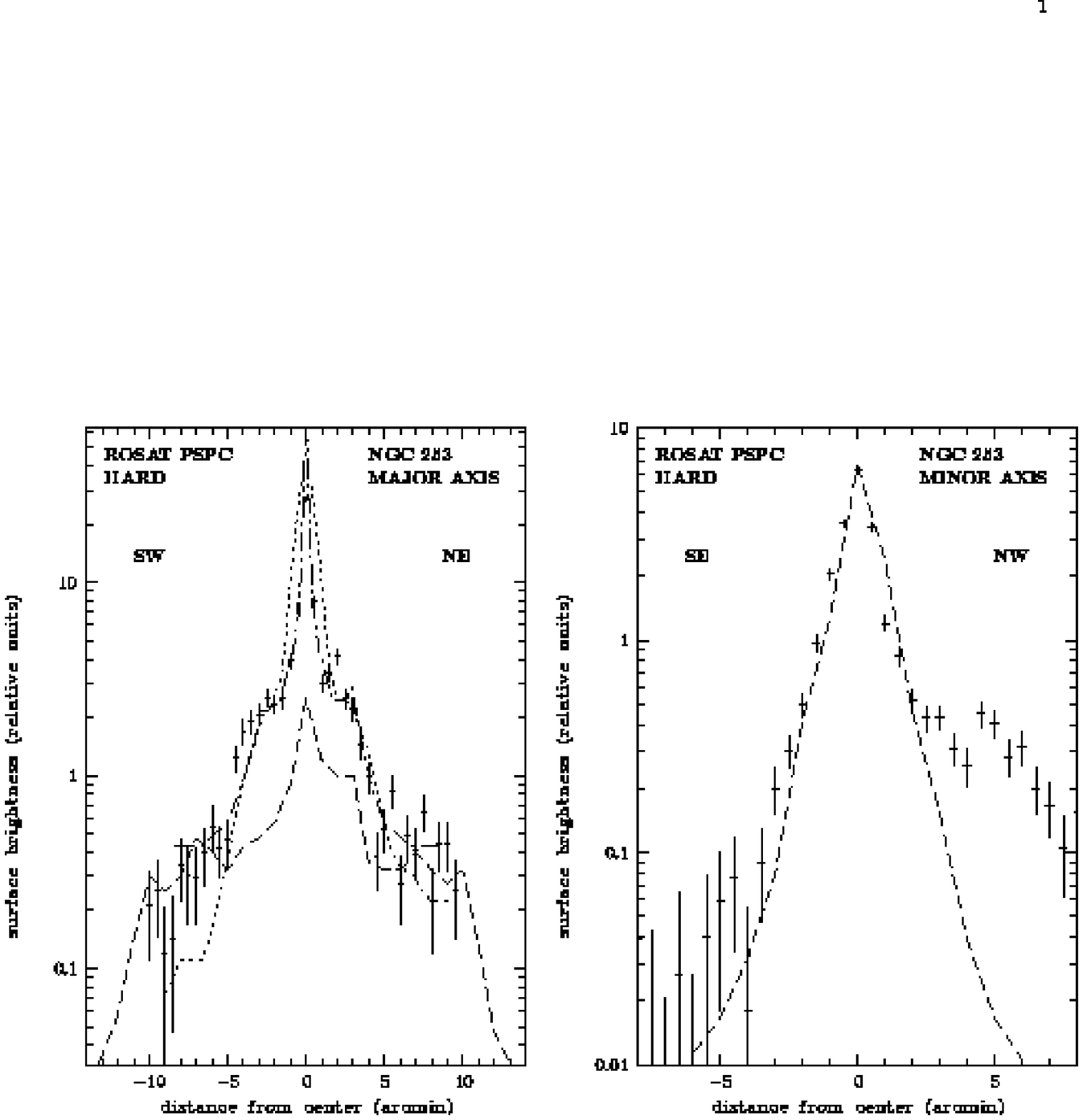}}
  \caption[]{
   Background corrected spatial distributions of diffuse surface brightness 
   along the major (left) and minor (right) axes of NGC~253 for ROSAT PSPC 
   hard band (0.5--2.0 keV, see 
   Fig. \ref{profile_ax}). Distance is given relative to the NGC 253 nucleus 
   with the SW disk region at left and NE to the right for the major axis
   and the SE halo region at left and NW to the right for the minor axis,
   respectively. Surface brightnesses from other wavelengths are added 
   (see text): dotted $\cor$ Radio 1.46 GHz profile 
   (Hummel et al. 1984), dashed $\cor$ optical (Pence 1980), dash dotted $\cor$ 
   infrared 2.2$\mu$m (Scoville et al. 1985), see text 
   }
   \label{ax_mult}
\end{figure*}
\begin{figure*}
 \resizebox{12cm}{!}{\includegraphics[bb= 57 120 399 733,clip=]{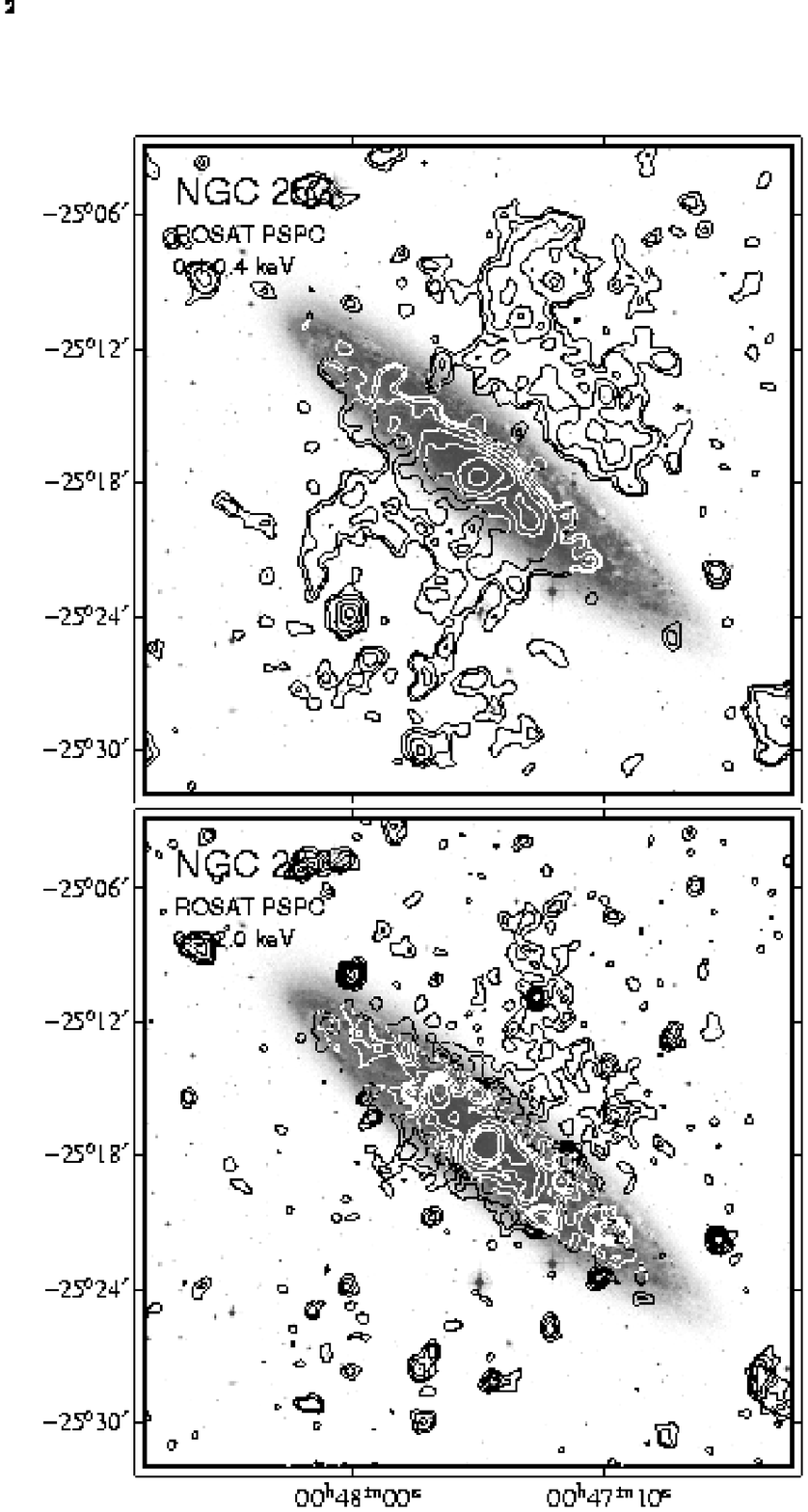}}
 \hfill
 \parbox[b]{55mm}{
  \caption[]{
   Contour plot for the ROSAT PSPC soft (0.1--0.4 keV, top) and
   hard band (0.5--2.0 keV, bottom) overlaid on an
   optical image of NGC~253 extracted from the digitized sky survey.
   X-ray contours are the same as in Fig.~\ref{four_in_one} for the
   soft band. For the hard band, contours are given in units of 1 
   photon accumulated per 28\arcsec\ diameter. One
   unit$\;=250\,10^{-6}$\, cts s$^{-1}$ arcmin$^{-2}$.
   Contour levels are 3, 5, 8, 12, 17, 30, 60, 120 units 
   }
   \label{opt}}
\end{figure*}

In Sect. 3.2.2, we 
identified at least three components to the diffuse emission of the NGC~253 
disk and the flat SE halo using the surface
brightness distributions along the major and minor axes 
(Fig. \ref{profile_ax}) 
that are seen to differ in spatial extent: 
\begin{enumerate}
\item a bright component that originates from the
nuclear region was discussed in the previous section and is connected to the 
extended source X34 and the X-ray plume 
\item a soft component most likely originating 
from hot gas immediately above the layer of the dense interstellar medium of 
the disk as can be inferred from the shift of its maximum to the SE by 
$\sim$45\arcsec. It rises to a maximum height of 2.2 kpc above the galaxy 
plane with
an extent of $\pm$4.5 kpc along the major axis (coronal emission above the 
disk) 
\item in the hard band, a bright 
central and a fainter extended component (extents of $\pm$3.4 kpc and 
$\pm$7.5 kpc, respectively) becomes visible  centered on the plane of
the galaxy and most likely originating from within the disk 
\end{enumerate}

\begin{figure}
  \resizebox{\hsize}{!}{\includegraphics[bb=70 10 440 535,angle=-90,clip=]{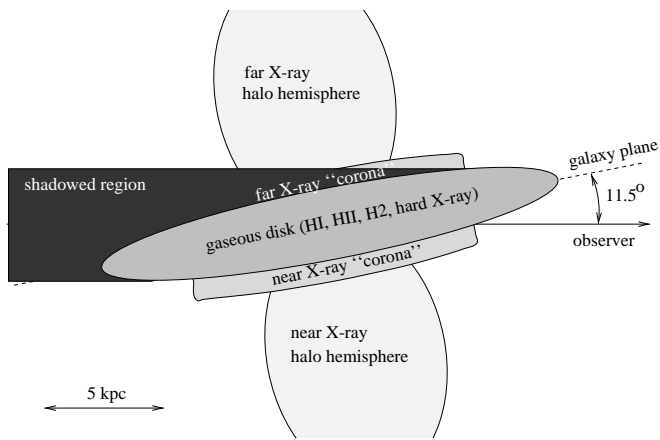}}
    \caption[]{
     Sketch of NGC 253 observing geometry and X-ray halo components
     }
    \label{sketch}
\end{figure}

Components 2 and 3 can clearly be 
identified in the overlays of the soft and hard PSPC X-ray contours on an 
optical image of NGC 253 (Fig. \ref{opt}). Component 2 may be further 
subdivided into emission connecting component 1 with the horn-like structure
in the outer halo hemisphere (as indicated by the curvature of the soft
band contours close to the nuclear area), reflecting the strong galactic wind
emanating from the starburst nucleus, and a component floating on the disk 
like a spectacle-glass. The latter most likely originates from hot gas 
fueled from galactic fountains (see below). 
The proposed location and viewing geometry of the
different halo components is sketched in Fig. \ref{sketch}.   

The presence of these bright
components hinders the detection of X-ray emission -- if present -- 
from the bar of NGC 253 
(seen e.g. in infrared data with position angle 70\degr, 
extent $\sim$150\arcsec; Forbes \& DePoy 1992). On the other hand, there 
is a clear correspondence in the brightness distribution of the diffuse 
hard band emission with the spiral structure in the inner and the 
outer disk of NGC 253 (Fig. \ref{opt}). South of the nucleus the inner 
spiral arm is located to the north and the brighter part of the outer 
spiral arm to the south within the projected elliptical shape of the disk.
North of the nucleus, the spiral arm moves outward from south to north (compare
optical and \HI\ analysis by Pence 1980 and Koribalski et al. 1995, 
respectively).

The X-ray surface brightness distributions (Fig. \ref{profile_ax}) 
can be compared to the emission at other wavelengths.
The soft band profile originates from above the disk, therefore 
one expects little similarities with emission at optical continuum and 
infrared wavelengths. The hard 
band, however, traces diffuse emission in the disk, and correlations with other
wavelengths are expected. Thus, for \ein IPC data Fabbiano (1988) reported  a 
good correlation with the radio profile along the major axis. In 
Fig. \ref{ax_mult} we compare the background 
corrected ROSAT PSPC hard-band profiles along the major and minor axes to 
optical, infrared, and radio profiles. 

The optical (blue light) profiles were extracted from Pence (1980, Fig. 8) 
and normalized to the wings of the major axis X-ray profile (7\farcm5 offset)
and to the maximum of the minor axis profile.    
Along the major axis the optical brightness shows a bright plateau out to a 
galacto-centric distance of 10\arcmin\ and coincides nicely with the faint
extended X-ray component. The bright central and the 
nuclear component seem less pronounced or totally missing, which may be
explained by extinction effects. Along the minor axis, X-ray and optical
brightness profiles are comparable within the disc. However, the optical data 
do not indicate halo emission, as can be most clearly followed in the X-ray
profiles in the NW.

Infrared profiles were published by Scoville et al. (1985). We have compared 
2.2 $\mu$m data (their Fig. 1) obtained with a 10\arcsec\ beam
from scans along the major axis, and normalized their profile to match our 
bright central component ($\sim$2\farcm5 offset). The NIR profile follows the 
X-ray brightness of the bright central component and of the wings quite well. 
The larger 
X-ray extent to the SW of the bright central component may be explained by
the differences in the infrared beam size and the X-ray slit length. As these 
systematic effects should be much stronger along the minor axis, we did not 
include these data in the figure. 
The excess 2.2 $\mu$m radiation, if compared with colours
normally seen in disk galaxies, cannot be entirely explained by heavy
extinction, but is most likely a contribution from clouds of hot dust 
associated with star formation regions.
 
While the radio 1.46 GHz radio emission (Hummel et al. 1984) emphasizes the
disk emission, the 330 MHz emission (Carilli et al. 1992, see Fig. \ref{radio})
discloses the true halo emission. As we are here comparing disk distributions,
we compare the 1.46 GHz profile (Hummel et al. 1984) collected with a
beam size of 68\arcsec$\times$36\arcsec. It is normalized to match the bright 
central component at the same offset as the infrared. 
A similar profile was reported by Klein et al. (1983) based on 10.7 GHz 
observations. While in 1.46 GHz, the nuclear component along the 
major axis is more extended than the X-ray or infrared emission, 
the central component is similar in width to the infrared. The extended wings, 
however, decline faster than X-rays and infrared. A detailed comparison 
of radio and X-ray properties and especially of the distribution along the
minor axis will be given in a separate paper (Ehle et al., in preparation).

How can we explain the similarities and the differences in the different 
wavelength regimes? The strong emission from the central source in all
wavelengths reflects the presence of the strong nuclear starburst. 
In addition, the bright inner disk emission in the radio, infrared and X-rays 
is most likely caused by enhanced star formation in this region. The 
star-forming regions, however, are located in \HII\ regions well within the 
disk and are most likely heavily obscured in the optical. This may explain 
why the bright central disk is not as obvious in this band compared to other
wavelengths. However, in a detailed analysis of the morphology of dark lanes
and filaments in the dust-rich disk, Sofue et al. (1994) identified 
arcs with heights of about 100 to 300 pc, connecting together two or more 
dark clouds, 
as well as loops and bubbles expanding into the disk-halo interface with
diameters of a few hundred pc to $\sim$1 kpc, and vertical dust filaments, 
almost perpendicular to the galactic plane and extending almost coherently
1 to 2 kpc into the halo. Sofue et al. propose a {\it boiling disk} model,
in which the filamentary structures develop due to star-forming activity in
the disk combined with the influence of magnetic fields. There is clearly 
enhanced activity in the inner disk. This picture of the boiling disk, 
with indications of outflow of hot gas into the low halo (corona of the disk), 
connects naturally with the relatively soft X-ray emission floating on top 
of the disk of the galaxy in the SE like a spectacle-glass. In galaxies
with starforming activity distributed over several areas in the disk, galactic
fountains may be the mechanism to fuel the halo with hot gas (see e.g.
the X-ray halo of NGC 891, Bregman \& Pildis 1994). In NGC 253, however, the 
superwind from the nuclear starbursts clearly dominates. 

Analyzing the X-ray spectra, a two-temperature thin thermal plasma model
was sufficient to describe the integral disk spectrum. The low-temperature 
component (temperature 0.2 keV, extra-Galactic luminosity 
$7.8\,10^{38}$ \ergsec) could be fitted assuming no absorption from within 
NGC 253, and can therefore be identified with the soft-band emission above 
the disk as deduced from the surface brightness profiles attributed to the
disk corona. The component with the higher temperature of 0.7
keV is heavily absorbed ($9.5\,10^{21}$ cm$^{-2}$ in addition to Galactic 
foreground) as expected if it originates from sources 
within the disk, and has an intrinsic luminosity of $1.2\,10^{39}$ \ergsec. 
The spectrum certainly represents a mixture of sources. Unresolved
point sources below the point source detection threshold (X-ray binaries,
SNRs) will contribute, as well as diffuse emission from \HII\ regions  and 
from the hot component of the interstellar medium. In principle, these 
components might be separated utilizing their characteristic spectra. However, 
due to the location in the disk, their spectra suffer from different levels 
of absorption so that 
observations with much better statistics and spatial and spectral 
resolution are required to resolve them.

Mean physical parameters for the flat SE halo component immediately above the 
disk can be inferred from the above results if we make some assumptions about
the geometry of the emission. Here we choose the simple geometry of a 
disk corona, an circular cylinder with a thickness of 1 kpc and a diameter of 
9 kpc with the central area cut out to a radius of 0.8 kpc. This approximates
the extraction 
area of the disk spectrum and the extent of the soft emission along the major 
axis. The results are not strongly dependent on these geometrical parameters.

\begin{table}
         \caption{Halo gas parameters for the different coronal components of
                  NGC 253 assuming thermal cooling and ionization equilibrium
                  for the diffuse components given in Table \ref{components}
                  and emission geometries as explained in Sect. 4.3}
         \label{halo_par}
         \begin{flushleft}
         \begin{tabular}{llll}
            \hline
            \noalign{\smallskip}
&$n_\mathrm{e}$&$m_\mathrm{gas}$&$\tau$\\
Region&(cm$^{-3}$)&(M$_{\sun})$&(y)\\
            \noalign{\smallskip}
            \hline
            \noalign{\smallskip}
SE disk corona&$1.4\,10^{-3}\,\eta_1^{-0.5}$
              &$2.7\,10^6\,\eta_1^{0.5}$
              &$1.6\,10^8\,\eta_1^{0.5}$\\
            \noalign{\smallskip}
NW halo soft &$5.8\,10^{-4}\,\eta_2^{-0.5}$
              &$7.6\,10^6\,\eta_2^{0.5}$
              &$2.0\,10^8\,\eta_2^{0.5}$\\
~~~~~~~~~~~~hard &$2.3\,10^{-4}\,\eta_2^{-0.5}$
              &$3.0\,10^6\,\eta_2^{0.5}$
              &$1.1\,10^9\,\eta_2^{0.5}$\\
            \noalign{\smallskip}
SE halo soft &$5.1\,10^{-4}\,\eta_3^{-0.5}$
              &$6.7\,10^6\,\eta_3^{0.5}$
              &$2.1\,10^8\,\eta_3^{0.5}$\\
~~~~~~~~~~~~hard &$1.2\,10^{-4}\,\eta_2^{-0.5}$
              &$1.6\,10^6\,\eta_2^{0.5}$
              &$2.0\,10^9\,\eta_2^{0.5}$\\
            \noalign{\smallskip}
            \hline
          \end{tabular}
         \end{flushleft}
   \end{table}

In this coronal model, the hot gas with the fitted temperature of 
0.2 keV $\cor 2.3\,10^6$ K (cooling coefficient of 
$2.4\,10^{-22}$ erg\,cm$^3$\,s$^{-1}$ according to Raymond et al. (1976)) 
and the unabsorbed luminosity of 
$7.8\,10^{38}$ \ergsec\ is distributed in a volume of $1.8\,10^{66}$ cm$^3$,
assuming a relative volume filling factor $\eta_1$ of the hot gas in clouds. 
Assuming thermal cooling and ionization
equilibrium (Nulsen et al. 1984), the electron density $n_{\rm e}$, 
mass $m_{\rm gas}$ and 
cooling time $\tau$ of the coronal gas are given in Table~\ref{halo_par}.
 
Neglecting effects of differing filling factors (which, however, would only 
contribute with the square root) the density of the soft and hard components 
in the outer halo hemispheres are lower 
by factors of 2 to 3 and more than 5, respectively,  than that of the disk 
coronal hot gas. 
The contributing mass in the hard components of the halo is similar to that in
the disk corona while it is higher by a factor of 2 to 3 in the soft halo 
components. The cooling time of the soft halo components is a factor of 2 
and that of the hard components more than a factor of 10 longer than  
in the disc corona. The scale height and properties of the coronal emission 
can be explained by galactic fountains (e.g. Breit\-schwerdt \& Komossa 1999 and 
references therein), created due to the heating of the hot intercloud medium by 
explosions of supernovae while the emission in the outer halo most likely 
originates from the galactic super-wind (e.g. Heckman et al. 1990, 
see Sect. 4.5). 

\subsection{Shadowing of halo emission by the disk}
An eye-catching structure in the ROSAT PSPC soft band image 
(Figs. \ref{four_in_one} and  \ref{opt}) and the surface
brightness profile along the minor axis (Fig. \ref{profile_ax}) 
is the depression/gap in the diffuse
emission between the galaxy disk and the NW halo hemisphere. The explanation
for this novel effect is straight forward. The diffuse emission from the 
corona and outer halo hemisphere on the far side of the galaxy is shadowed 
by the dense interstellar medium of the intervening
disk. In addition, in the gap area no emission from the other hemisphere 
contributes, due to the near edge-on view (see Fig. \ref{sketch}). 
Therefore this shadowing effect 
not only unequivocally reveals the geometry of the system (i.e. the NW edge
is the {\it near} side of NGC 253), but also allows us to determine lower 
limits to the column density of the intervening material.

\begin{table}
         \caption{Simulated reduction factors for the ROSAT PSPC count rate 
                  in the soft
                  (0.1--0.4 keV) and hard (0.5--2.0 keV) band for
                  thin thermal plasma spectra of temperature T and an 
                  additional absorbing column N$_{\rm H}$. We compare 
                  simulated count rates for thin thermal plasma spectra
                  (THPL) of temperature T  and total absorption 
                  N$_{\rm H} + $ N$_{\rm Hgal}$ to count rates for the same
                  THPL model and just Galactic foreground absorption 
                  (N$_{\rm Hgal}$). Morrison \&McCammon (1983) absorption
                  cross setcions are used}
         \label{abs}
         \begin{flushleft}
         \begin{tabular}{rrrrrrr}
            \hline
            \noalign{\smallskip}
N$_{\rm H}$&\multicolumn{2}{c}{T = 0.2 keV}&\multicolumn{2}{c}{T = 0.3 keV}
           &\multicolumn{2}{c}{T = 0.4 keV}\\
&soft&hard&soft&hard&soft&hard \\
(10$^{20}$ cm$^{-2}$)\\
            \noalign{\smallskip}
            \hline
            \noalign{\smallskip}
2.5 &  2.6&1.2& 2.6&1.1& 2.6&1.1\\
5   &  5.1&1.4& 5.1&1.3& 5.2&1.2\\
10  &   17&1.9&  16&1.6&  16&1.5\\
20  &  120&3.4& 100&2.6& 100&2.3\\
40  & 3400& 10&2700&6.1&2900&4.8\\
            \noalign{\smallskip}
            \hline
          \end{tabular}
         \end{flushleft}
   \end{table}

To estimate the effects of absorption, we could assume for the unabsorbed
flux in the NW an identical profile as observed in the SE and 
calculate the amount of absorption from the reduction factor determined 
from the profile really measured. However, we already pointed out in 
Sect. 3.2.2 that the fluxes in the unabsorbed parts in the NW  
are $\sim$1.5 times brighter than in the SE at similar nuclear distances. 
Also, the spectrum in the outer halo is significantly harder. 
In Table \ref{abs} we put together reduction factors with respect to pure 
Galactic foreground absorption for the 0.1--0.4 keV and 0.5--2.0 keV ROSAT 
PSPC bands, calculated for thin thermal plasma spectra, with 
temperatures of 0.2, 0.3 and 0.4 keV and 
increasing additional shadowing columns of cold gas (N$_{\rm H}$ of 
2.5 to $40\times10^{20}$ cm$^{-2}$). While, for a given absorbing column, the 
reduction factors for soft band emission do not vary strongly for the 
temperature range investigated, for a fixed temperature there is strong 
variation with increasing absorption (factors of $\le$1.2 and $>1000$, 
respectively). The corresponding hard band variations are stronger for 
changing temperatures, but much weaker with absorption (2 and 10, respectively). 

\begin{table}
         \caption{Approximate measured reduction factors for the ROSAT PSPC 
                 count rate in the soft (0.1--0.4 keV) and hard (0.5--2.0 keV) 
                 band, determined for different major axis offset angles 
                 to the NW as described in Sect. 4.4 based on the minor axis 
                 profiles (Fig. \ref{profile_ax})}
         \label{red}
         \begin{flushleft}
         \begin{tabular}{lrrrr}
            \hline
            \noalign{\smallskip}
Offset&1\arcmin&2\arcmin&3\arcmin&4\arcmin \\
            \noalign{\smallskip}
            \hline
            \noalign{\smallskip}
Soft &$>70$&$>50$&4&2 \\
Hard &8&6&3&2 \\
            \noalign{\smallskip}
            \hline
          \end{tabular}
         \end{flushleft}
\end{table}

The theoretically expected reduction factors can be compared to factors 
determined from the minor axis profiles (Fig. \ref{profile_ax}), by dividing 
count rates extrapolated from larger off-axis angles to the inner disk by
the measured count rates (Table \ref{red}). At an off-axis angle of 4\arcmin\ 
the factors still indicate absorbing columns of $\sim2\,10^{20}$ \cm-2
and at 3\arcmin\ of $4\,10^{20}$ \cm-2 using the soft-band information. The
less reliable extrapolation of the hard band would indicate even higher 
absorption.

These results can be compared with \HI\ measurements. Puche et al. (1991) present
data obtained with a beam of 68\arcsec\ diameter. Their lowest contour of 
$2.4\,10^{20}$ \cm-2 is offset from the nucleus along the minor axis to 
the NW by 3\arcmin. With a beam diameter of 30\arcsec\ the lowest contour of 
$4.8\,10^{20}$ \cm-2 in the map of Koribalski et al. (1995) shows the same 
offset. As explained in detail in Sect. 4.2.1, \NH\ is determined by 
contributions from \HI\ and H$_2$. The H$_2$ distribution of NGC 253 can
be inferred from CO maps (Houghton et al. 1997). While these maps have less 
resolution than the \HI\ maps, they more or less coincide in general extent
and imply a significant contribution to \NH.
Using our X-ray absorption technique, we can determine absorption 
columns in the NW outer disk with similar accuracy. Unfortunately, the X-ray
shadowing method is not generally applicable to measure the density of the
interstellar medium in galaxies, but NGC 253 reflects a case of special luck.  
Galactic foreground \NH\ needs to be rather low ($\la 1.5\,
10^{20}$ \cm-2), the galaxy inclination has to be close to edge on, and -- 
the most important requirement -- the galaxy halo must emit soft X-rays to 
allow this kind of study. 

Even for our NGC 253 analysis one has to keep in mind that the count
rates are averaged along the disk at varying absorption depths and
X-ray brightnesses. Data with better statistics as expected from the upcoming 
XMM-Newton observatory will allow the determination of the absorbing column along 
the NW outer disk with good
spatial resolution. Such data should also clearly uncover the geometry of the 
outflow to the NW. Its soft emission is hidden by the disk, however it shines
through at energies above 0.5 keV (see the hard1 and hard2 images in 
Fig. \ref{four_in_one}), indicating that
unfortunately the hemisphere with the more spectacular outflow is shadowed 
by the NGC 253 disk.

\subsection{Extended soft X-ray halo of NGC~253}
\begin{figure*}
 \resizebox{12cm}{!}{\includegraphics[bb= 55 255 506 671,clip=]{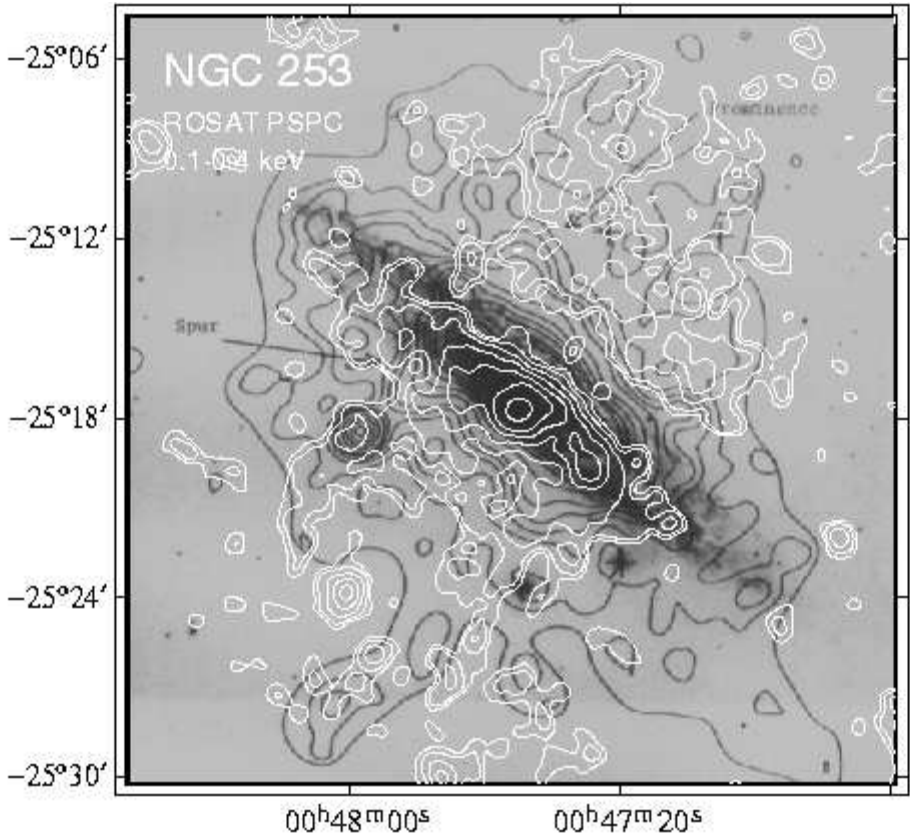}}
 \hfill
 \parbox[b]{55mm}{
  \caption[]{
   Contour plot for ROSAT PSPC soft band (0.1--0.4 keV) overlaid in white on 
   a radio map at 330 MHz (black contours over an optical image (Fig. 2 from
   Carilli et al. 1992).
   X-ray contours are the same as in Fig.~\ref{four_in_one}
   }
   \label{radio}}
\end{figure*}

The detection of hot gas in the halo of NGC 253 extending to projected 
distances of 9 kpc from the galactic plain, allows us to investigate a 
component of the interstellar medium that cannot be accessed in optical 
observations. The X-ray structure (Sect. 3.2.2) clearly 
indicates that the emission is not filling the halo homogeneously. Instead,
the soft image resembles a horn-like structure in both hemispheres, as would be 
expected if the emission regions were shaped like a hollow-cone. While 
the hemispheres look rather similar in the broad band image, they differ 
significantly in the hard band. This behavior is also reflected in the 
spectral parameters, the NW halo hemisphere being significantly hotter than 
the SE one. In addition to this difference between the two halo 
hemispheres, the northern most ``horn" in both hemispheres is more pronounced.

Analogously to the corona above the disk (see Sect. 4.3), we can determine mean
physical properties for the emitting medium. For simplicity we assume that the 
hot gas in both hemispheres is filling a cylinder of 4.5 kpc radius and 
7 kpc height. It is evident from the images, that the filling factors $\eta_2$ and
$\eta_3$ for the halo hemispheres differ. 
As discussed in Sect. 3.2.2, no simple model results in a good 
approximation for the individual 
halo spectra. Nevertheless, to derive halo gas parameters 
(cf. Table \ref{halo_par}), we made use of the temperatures  
(0.19 keV $\cor  2.2\,10^6$ K and 0.12 keV $\cor  1.4\,10^6$ K)
and corresponding unabsorbed luminosities  
($1.0\,10^{39}$ \ergsec\ and $5\,10^{38}$ \ergsec), 
derived from a thin thermal plasma model fit to the NW and SE 
halo hemispheres, respectively. The significantly higher luminosity and 
temperature in the NW with respect to the SE halo hemisphere nearly cancel 
each other and only 
lead to slightly higher electron densities and masses, while the cooling
time stays the same.

The ROSAT halo parameters in the NW are consistent with the parameters derived 
from \ein observations (Fabbiano 1988), once corrected to our assumed
distance and to the ROSAT measured temperature. The sum of our components 
differs, however, significantly from the parameters derived by Read et al. 
(1997) using integral properties of the diffuse emission of NGC 253; they find
lower gas masses and shorter cooling times as well as allowing for lower 
electron densities in the radiating plasma.

If the hot gas in the outer halo represents nuclear and disk material 
transported into the halo by the super-wind 
with a velocity of $\sim$350 km s$^{-1}$ as modeled for the 
optical emission cone close to the nucleus (cf. Sect. 4.2.2), we can calculate
from the halo extent of 9 kpc a lower limit for the age of the wind 
of $2.5\,10^7$ y. This time is comparable to the
age of the rapid massive star formation in the nucleus of NGC 253, derived
from evolutionary synthesis models ($2-3\times10^7$ y, e.g. Engelbracht 
et al. 1998), it is shorter, however, by at least an order of magnitude than
the cooling time of the hot gas in the halo calculated above. From these
considerations it is possible that the hot halo gas was heated in
the starburst nucleus and transported into the halo by the wind. However, one 
could also consider that the wind is heating ambient gas in the halo. 

The first scenario has been proposed by Tomisaka \& Ikeuchi (1988). Their 
model of a galactic scale bipolar flow assumes a disk component to represent the
interstellar medium of the starburst galaxy, which for some time confines  
the hot matter within a cool shell, which is further heated by supernova 
explosions of massive stars formed in the nuclear starburst. Eventually, the 
shell breaks up and releases the hot gas into the halo. Suchkov et al. (1994) 
improved the model by including an ambient two-component disk-halo interstellar
medium and argued that this two-component representation is crucial for
adequate modeling of starbursts. In their models, the bulk of the soft thermal
X-ray emission from starbursts arises in the wind-shocked material of the
halo and disk gas transported into the halo, rather than in the hot wind 
material itself. The super-wind itself is too hot and thin, to be visible in 
the ROSAT band and reaches speeds of $\sim$2000 km s$^{-1}$. The models allow 
to produce X-ray geometries and filamentary structures similar to those 
observed in NGC 253 and propose a temperature distribution  of the heated 
material of $2-5\times10^6$ K. However, they neglect the role of cosmic rays, 
magnetic fields, and delayed recombination (Breit\-schwerdt \& Komossa 1999).      

The spectra from the halo hemispheres cannot be fitted by simple thin thermal plasma 
models. However, two temperature models or 
a one temperature model with additional emission components 
above 0.7 keV give reasonable fits to the data (Sect. 3.2.2). This
confirms on a smaller scale results by e.g. Dahlem et al. 1998 and 
Strickland \& Stevens 1998 who found observational evidence
that the spectra of the diffuse soft X-ray emission in galactic winds or 
superbubbles suggest more than a single temperature. The temperatures
and luminosities of the observed NGC 253 halo components are
consistent with luminosities and the temperature distribution proposed 
by the models for the heated material in the halo.

Breit\-schwerdt \& Schmutzler (1999) showed that the assumption
of cooling via collisional ionization equilibrium used in the thin 
thermal plasma models are not correct if the dynamical time scale of the 
plasma, as for instance in an out-flowing wind, is much shorter than the 
intrinsic time scales (e.g. recombination, collisional excitation,
ionization etc.). Therefore, the dynamical and thermal evolution of the
plasma has to be treated self-consistently. First results show that the
expected spectra of cooling wind material in the halo should exhibit enhanced 
emission in lines at energies above $\sim$0.7 keV (e.g. from O\,{\sc vii},
O\,{\sc viii},
and highly ionized Fe), compared to equilibrium models, as well as a lower 
plasma temperature due to adiabatic expansion. Since recombination of highly 
ionized species is delayed, the resulting X-ray line emission reflects the
``initial temperature" of the plasma close to the disk, when it was much 
hotter. In these non-equilibrium models one cannot determine a single
cooling timescale, ``equilibrium cooling" due to expansion is likely to be 
much shorter than the equilibrium value given in Table \ref{halo_par} and
the time scale for line recombination longer. Numerical simulations for
a galactic outflow including non-equilibrium X-ray emission in a 
self-consistent fashion have been folded through the ROSAT PSPC instrumental 
response (Breitschwerdt \& Freyberg 2000). Best fitting spectral models
are two temperature thin thermal plasma or a one temperature model with an
additonal Gaussian component similar to our finding. Breitschwerdt \& Freyberg 
argue that the two fitted temperatures are not physical but mimic 
non-equilibrium X-ray emission. 
  
Keeping in mind the super-wind origin of the halo emission, one can consider 
several explanations for the asymmetries in the halo X-ray emission. The
asymmetry may either be caused by differing wind parameters in both 
hemispheres, originating for instance from a slight asymmetry of the 
driving starburst, or by a different structure of the ambient halo medium
MacLow \& McCray 1988). These explanations may well explain the asymmetry 
between the SE and NW halo, they also may account for the stronger ``horns" 
in the north than in the south of both hemispheres. One can, however, think
of other possibilities to generate the asymmetry of the ``horns".
There could be more ambient material on the N side of the galaxy to 
shock-exite (perhaps due to an earlier fountain episode?), or even the 
interaction of the halo gas due to  the relative motion with 
the intergalactic medium of the Sculptor group of galaxies, of which 
NGC 253 is a member (e.g. Puche \& Carignan 1988), could be responsible. 
Unfortunately, neither the 
inter-galaxy medium nor the relative motions are known for the Sculptor 
group galaxies to strengthen or reject the latter hypothesis.

The X-ray halo has no equivalent in the optical continuum. The optical 
continuum halo
mainly follows the elliptical shape of the galactic disk to distances along the
major axis of 22.5 kpc (Pence 1980, Beck et al. 1982) and is most likely made 
up of late type stars. This situation is different for the H$\alpha$ emission. 

Radio images at 
330 MHz (Carilli et al. 1992) show a synchrotron emitting halo which  
is more extended perpendicular to the disk and in general matches the 
soft X-ray halo (cf. Fig. \ref{radio}). At radio frequencies of
1.5 GHz, the emission is box-shaped similar to the disk corona emission
(Hummel et al. 1984, Carilli et al. 1992, Beck et al. 1994).
Carilli et al. present evidence for outflow from the disk in a "spur" which
-- corrected for the NGC 253 distance assumed in this paper -- rises 2.5 kpc
above the plane at about the position of the NE X-ray horn. Detailed comparison 
of the morphology of the halo emission is hampered by the poorer resolution 
of the 0.33 GHz radio (60\arcsec\ and outermost contour 120\arcsec) 
with respect to the soft X-ray contours (40\arcsec). However, there is 
indication from the overlay
that the radio emission does not show the horn-like structure which dominates
the X-ray morphology. So in X-rays, the curvature of the NE horn points at
a connection to the nucleus, while the NE radio spur seems to be directly 
connected to the disk and offset to the north. Radio and X-ray data with 
better resolution and statistics are urgently needed to clarify these
differences, that can not be understood in terms of models put forward
for the common origin of the X-ray and radio halo emission.
For the NGC 253 radio halo, Beck et al. (1994) extracted magnetic 
fields and rotation measures using multi-frequency observations of the radio 
continuum emission. X-ray, radio and H$\alpha$ measurements will be compared 
in detail in a separate paper (Ehle et al., in preparation) 
to determine the importance of magnetic and thermal 
effects for the interstellar medium in the galaxy disk and halo.

\subsection{Comparison with diffuse X-ray emission detected in other 
spiral galaxies}
   \begin{table}
      \caption{Halo parameters of nearby edge-on starburst galaxies:
               for the galaxies the extent of the X-ray halo perpendicular
               to the disk (z), the average temperature (T) and the  
               luminosity (L$_{\rm X}$) are given as well as the approximate 
               mass of the hot halo gas (m$_{\rm gas}$)}
         \label{starburst}
         \begin{flushleft}
         \begin{tabular}{lrrrrr}
            \hline
            \noalign{\smallskip}
Galaxy &z &T & L$_{\rm X}$ &$\eta^{0.5}\cdot m_{\rm gas}$&Ref.  \\
       &(kpc)&(keV)&($^\ast$)&($10^7$ M$_{\sun}$)& \\
            \noalign{\smallskip}
            \hline
            \noalign{\smallskip}
NGC 253 & 9 & 0.15 & $2$&$2$& (1)\\
M82     & 6 & 0.5  & $19$ &$13$& (2) \\
NGC 3079& 13.5 & 0.3& $6$ &$20$& (3) \\
NGC 3628& 25 & 0.16 & $8.3$ &$20$& (4) \\
            \noalign{\smallskip}
            \hline
            \noalign{\smallskip}
         \end{tabular}
         \end{flushleft}
{
$^\ast$:  in units of $10^{39}$ \ergsec \\
References: (1) this work; (2) Strickland et al. 1997; (3) Pietsch et al. 1998; 
(4) Dahlem et al. 1996
}
   \end{table}

With the help of the ROSAT satellite, diffuse emission from the disk and halo 
of several late-type spiral galaxies was separated, irrespective of their 
face-on or edge-on orientation. References for 
some galaxies analyzed in detail and used for comparison to our NGC 253 
results have already been given or are given below. In addition, 
Read et al. (1997) 
present a homogeneous, however less deep, analysis of 17 nearby galaxies 
partly overlapping with our sample, and determine rough parameters for 
the diffuse emission. Detailed analysis of close to face-on galaxies 
(e.g. M 101, Snowden \& Pietsch 1995; NGC 4258, Pietsch et al. 1994; 
NGC 4449, Vogler \& Pietsch 1997; M 83, Ehle et al. 1998; NGC 7793, Read \&
Pietsch 1999) clearly show
diffuse emission. The separation in disk and halo components proved difficult
and was mainly based on the absorbing columns derived from spectral fits.
The separation is much easier for (close to) edge-on galaxies due to the fact 
that in these systems disk and extended halo components spatially separate. 
Diffuse emission has not been detected from the halo of all ``normal" edge-on 
galaxies. We list some examples with halo detection indicated in brackets:
NGC 4565 (+), NGC 4656 (?), NGC 5907 (-) (Vogler et al. 1996); 
NGC 4631 (+) (Wang et al. 1995, Vogler \& Pietsch 1996); 
NGC 4559 (-) (Vogler et al. 1997). However, the galaxies showing  X-ray 
emission from the halo most convincingly, are really the nearby 
edge-on starburst galaxies like NGC 253, M82, NGC 3079, and NGC 3628.

While the halo of NGC 253 presents a spectacular image and clearly 
stands out compared to other normal galaxies, its properties do not at all 
beat other nearby edge-on starburst galaxies. In
Table \ref{starburst} we compare the z extent of the halo above the disk as 
well as the temperature T, absorption corrected luminosity (0.1-2.4 keV) 
L$_{\rm X}$ of the halo gas and inferred gas mass $m_{\rm gas}$. From Table 
\ref{starburst} one would gather that in view of the halo extent, NGC 253 
is not on the low side of the sample. Recently, however, Lehnert et al. 
(1999) reported  emission at a distance of 11.6 kpc for the smaller M82 halo, 
which they interpret as being due to shock-heating of a massive
ionized cloud by the starburst super-wind, which would clearly exceed 
the NGC 253 halo extent. 
Apart from NGC 3628, the temperatures 
of the halo gas of the other galaxies are higher by a factor of 2 to 3
and the luminosities by up to a factor of 10 implying gas masses larger by
up to a factor of 20. Therefore, it is not the extreme halo 
properties of NGC 253, that have pushed it to the front of galaxy halo 
investigations, but its proximity and low Galactic foreground absorption.
It is interesting to note that for the four galaxies in the above sample,
besides starburst activity, LINER activity has also been discussed, pointing 
to the presence of an active nucleus. All four galaxies are members of dense
groups, and at least for M82, NGC 3079 and NGC 3628 it is clear that 
interactions within the group have most likely initiated the starburst (and 
AGN) activity. For NGC 253, we cannot be sure that a recent 
encounter with neighboring galaxies of the Sculptor group has 
initiated the activity. As an alternative explanation for the starburst, 
bar activity has been put forward. However again, a `flyby' of another 
Sculptor group galaxy might have caused the bar, which in turn may have 
caused the starburst. 

\section{Summary and conclusions}
Following the detailed analysis of deep ROSAT HRI and PSPC observations 
for X-ray point sources (Paper I), we here have characterized the diffuse X-ray 
emission of this edge-on starburst galaxy. After subtraction of 
the point-source contribution we determined the geometry and spectra of the
diffuse components and discuss our results in view of observations at other
wavelengths and from other galaxies. In detail, X-ray, radio, and
H$\alpha$ measurements will be compared in a separate paper
(Ehle et al., in preparation). Our main results can be summarized as follows:

\begin{itemize}
\item The diffuse X-ray luminosity of NGC 253 is distributed in about 
equal emission components from nuclear area, disk, and halo and contributes 
$\sim$80\% to the total X-ray luminosity of NGC 253 
(L$_{\rm X} = 5\,10^{39}$ \ergsec, corrected for foreground absorption). The
starburst nucleus itself is highly absorbed and not visible in the ROSAT band.
\item The ``nuclear source" (X34) has an extent of 250 pc (FWHM) and is located 
about 100 pc above the nucleus along the minor axis on the near side of 
NGC 253. It is best described as having a thermal
bremsstrahlung spectrum with a temperature of T = 1.2 keV
(N$_{\rm H} = 3\,10^{21}$ cm$^{-2}$) and
L$_{\rm X}^{\rm exgal} = 3\,10^{38}$ \ergsec\
(corrected for Galactic foreground absorption). The nuclear source most likely 
marks the area where the line-of-sight absorption gets low enough 
that soft X-ray emission from the heated walls of the funnel drilled by the 
super-wind, can leak through.
\item The ``X-ray plume" has a hollow-cone shape (opening angle of 32\degr\ and
extent of $\sim$ 700 pc along the SE minor axis). Its spectrum is best modeled 
by a composite of a thermal bremsstrahlung
(N$_{\rm H} = 3\,10^{20}$cm$^{-2}$, T = 1.2 keV, L$_{\rm X}^{\rm exgal} =
4.6\,10^{38}$ \ergsec) and a thin 
thermal plasma (Galactic foreground absorption, T = 0.33 keV,
L$_{\rm X}^{\rm exgal} = 4\,10^{38}$ \ergsec). It traces the interaction region
between the galactic super-wind and the dense interstellar medium of the disk. The
soft component with just Galactic foreground absorption stems from the halo 
above the disk.
\item Diffuse emission from the disk is heavily absorbed and follows the spiral
structure. It can be described by a thin thermal plasma spectrum ( T = 0.7 keV,
intrinsic luminosity L$_{\rm X}^{\rm intr} = 1.2\,10^{39}$ \ergsec),
and most likely reflects a mixture of sources (X-ray binaries, supernova
remnants, and emission from \HII\ regions) and the hot interstellar medium.
The surface brightness profile, with a bright inner and a fainter outer
component along the major axis with extents of $\pm$3.4 kpc and $\pm$7.5 kpc,
resembles the 1.46 GHz radio profile.
\item The ``coronal emission" originates from the halo immediately above the 
disk (scale height $\sim1$ kpc). It is only detected from the
near side of the disk (in the SE, T = 0.2 keV,
L$_{\rm X}^{\rm intr} = 7.8\,10^{38}$ \ergsec), 
emission from the back (in the NW) is shadowed by the intervening 
interstellar medium causing a pronounced gap in the soft X-ray emission 
along the NW edge of the optical disk, unambiguously determining the
orientation of NGC 253 in space. The X-ray corona is primarily due to
the nuclear superwind that is also responsible
for the emission in the outer halo. An additional component may be due to
hot gas ejected by galactic fountains originating within the boiling 
star-forming disk. 
\item The emission in the outer halo
can be traced to projected distances from the disk of 9 kpc, and shows a
horn-like structure. Luminosities are higher (10 and $5\,10^{38}$ \ergsec, 
respectively) and spectra harder in the NW halo than in the SE. 
The emission most likely originates from a strong galactic wind emanating
from the starburst nucleus. A two temperature thermal plasma model
with temperatures of 0.13 and 0.62 keV or a thin 
thermal plasma model with temperature of 0.15 keV and Gaussian
components above $\sim$0.7 keV and Galactic foreground absorption 
are needed to arrive at acceptable fits for the NW halo. This
may be explained by starburst-driven super-winds or by effects of a 
non-equilibrium cooling function in a plasma expanding in a fountain or wind.
\end{itemize}

While NGC 253 does not beat other nearby edge-on starburst galaxies with its 
X-ray halo parameters, it stands out due to the low foreground \NH,
distance, and the favorable inclination, which all together allowed the 
detection of so many details with ROSAT. Of specific interest are three
questions that only could be touched but not finally answered by ROSAT and
ASCA/BeppoSAX:
\begin{enumerate}
\item Is the X-ray emission in the outer halo caused by the nuclear super-wind 
or by hot gas transported into the halo via fountains that are fed from the 
star forming, boiling disk?
\item Can the X-ray spectra of the halo really be described by a delayed 
recombination model? 
\item Is the nuclear starburst region responsible for the nuclear and X-ray 
plume emission or is there a contribution from jet-like activity of a hidden AGN?
\end{enumerate}
These and more questions should be solved with the next generation of X-ray 
instruments on board XMM-Newton and \chandra, that will have a broader energy coverage,
higher collecting area, and much better spatial and energy resolution. 
Therefore, NGC 253 promises to stay a key object to understand the formation
and fueling of hot galaxy halos with future X-ray observations.

\begin{acknowledgements}
Hartmut Schulz kindly provided us with the H$\alpha$ images for the HRI 
overlays of the central field. We are very grateful to Dieter Breitschwerdt 
for discussions of the effects
of non-equilibrium cooling, and Andy Read and Ginevra Trinchieri for a
careful reading of the manuscript. This research has made use of 
the NASA/IPAC Extragalactic Database (NED) which is operated by
the Jet Propulsion Laboratory, CALTECH, under contract with the National
Aeronautics and Space Administration. To overlay the X-ray data we used an image
based on photographic data of the National Geographic Society -- Palomar
Observatory Sky Survey (NGS-POSS), obtained using the Oschin Telescope on
Palomar Mountain.  The NGS-POSS was funded by a grant from the National
Geographic Society to the California Institute of Technology.  The
plates were processed to the present compressed digital form with
their permission.  The Digitized Sky Survey was produced at the Space
Telescope Science Institute under US Government grant NAG W-2166.
The ROSAT project is supported by the German Bundesministerium f\"ur
Bildung und Forschung (BMBF/DLR) and the Max-Planck-Gesellschaft (MPG).
\end{acknowledgements}

\end{document}